\documentclass[a4paper,oneside,12pt]{article}

\usepackage{float,epsfig,array,amsmath,amssymb,amsfonts,latexsym,psfrag,graphicx,bm}
\usepackage{color,comment}
\usepackage{cite}

\newcommand{\nn}{\nonumber}
\newcommand{\Eqn}[1]{&\hspace{-0.5em}#1\hspace{-0.5em}&}
\newcommand{\be}         {  \begin{equation}  }
\newcommand{\ee}           {  \end{equation}  }
\newcommand{\eqb}{\begin{eqnarray}}
\newcommand{\eqe}{\end{eqnarray}}
\def\ba  { \begin{align} }
\def\ea  { \end{align} }

\def\ep{\epsilon}

\def\o{\over}
\def\cdots {\cdot\cdot\cdot}
\def\pint#1 {- \!\!\!\!\!\!\!\! \,\int_{#1}}

\def\comma      { \, , }
\def\period     { \, . }

\def\abs#1      {  \, \vert #1 \vert \,   }
\def\binom#1#2 { \vecii{ {}_{#1} }{\raisebox{.5ex}{$ {}^{#2} $}} }

\def\blam    { \bm{\lambda} }

\DeclareMathOperator{\Tr}{Tr}

\DeclareMathOperator{\diag}{diag}

\newcommand{\bbR}{{\mathbb R}}
\newcommand{\bbZ}{{\mathbb Z}}

\newcommand{\mf}[1]{\mathfrak{#1}}

\def\papertitlepage{\baselineskip 3.5ex \thispagestyle{empty}}
\def\preprintnumber#1#2{\hfill
\begin{minipage}{1.2in}
#1 \par\noindent #2 
\end{minipage}}
\renewcommand{\thefootnote}{\fnsymbol{footnote}}

\def\tilal { \tilde{\alpha} }
\def\tilL { \tilde{L} }
\def\hatp { \hat{p} }
\def\calT {{\cal T}}
\def\qbar {{\bar{q}}}

\def\TTheta {{\vartheta}}

\makeatletter \@addtoreset{equation}{section} \makeatother
\renewcommand{\theequation}{\arabic{section}.\arabic{equation}}

\newcommand{\al}{\alpha}

\newcommand{\La}{\Lambda}

\newcommand{\msc}[1]{\mbox{\scriptsize #1}}
\newcommand{\dsp}{\displaystyle}

\newcommand{\bz}{\Bbb Z}

\renewcommand{\-}{{\bf -1}}

\newcommand{\cJ}{{\cal J}}
\newcommand{\cT}{{\cal T}}

\newcommand{\tL}{\tilde{L}}

\newcommand{\tchi}{\tilde{\chi}}

\renewcommand{\th}{{\theta}}

\renewcommand{\mod}{\mbox{mod}}

\renewcommand{\nn}{\nonumber\\}

\newcommand{\tNS}{\widetilde{\mbox{NS}}}

\newcommand{\sNS}{\msc{NS}}

\newcommand{\sR}{\msc{R}}

\newcommand{\any}{{}^{\forall}}

\newcommand {\eqn}[1]{(\ref{#1})}

\begin{document}
\papertitlepage
\setcounter{page}{0}

\preprintnumber{UTHEP-695}{}

\vskip 8ex

\baselineskip=4ex
\begin{center}
{\Large\bf\mathversion{bold}
Lie algebra lattices and strings on T-folds
}
\end{center}

\vskip 0.5 ex 
\baselineskip=3.6ex
\begin{center}
  Yuji Satoh\footnote[1]{\tt ysatoh@het.ph.tsukuba.ac.jp} 
  and
  Yuji Sugawara\footnote[2]{\tt ysugawa@se.ritsumei.ac.jp}
 
\vskip 4ex
$^\ast${\it Institute of Physics, University of Tsukuba}\\
 {\it Ibaraki 305-8571, Japan}\\

\vskip 3ex
$^\dagger${\it Department of Physical Sciences, College of Science and Engineering}\\
 {\it Ritsumeikan University, Shiga 525-8577, Japan}
 
\end{center}

\baselineskip=3.6ex
\vskip 8ex
\begin{center} {\bf Abstract} \end{center}
\vskip 2ex
We study the world-sheet conformal field theories for T-folds
systematically based on the Lie algebra lattices representing the momenta of strings. 
The fixed point condition required for the T-duality twist  restricts the possible 
Lie algebras. When the T-duality acts as a simple chiral reflection, one is left 
with the four cases,  $A_1, D_{2r}, E_7, E_8$, among the simple simply-laced algebras.
From the corresponding Englert-Neveu lattices, 
we construct the modular invariant partition functions for the T-fold CFTs
in bosonic string theory. Similar construction is possible also by using Euclidean 
even self-dual lattices. We  then apply our formulation 
to the T-folds in the $E_8 \times E_8$
heterotic string theory. Incorporating non-trivial phases for the T-duality twist,
we obtain, as simple examples, a class of modular invariant partition functions
parametrized by three integers.  Our construction includes 
the cases which are not reduced to the free fermion construction.

\vskip 3ex
\vspace*{\fill}
\noindent
November, 2016
%
%

\setlength{\parskip}{1ex}
\renewcommand{\thefootnote}{\arabic{footnote}}
%
\setcounter{footnote}{0}
\setcounter{section}{0}
\pagestyle{plain}

\baselineskip=3.6ex
\newpage

\section{Introduction}

A salient feature of string theory is that physics on different background
geometries can be equivalent due to duality symmetries. 
This allows us to  think of geometries whose coordinate patches
are glued by duality transformations, as well as by ordinary general coordinate 
transformations \cite{Dabholkar:2002sy,Hellerman:2002ax,Flournoy:2004vn}.
They are relevant in understanding the vacua and the symmetries of string theory,
and may give clues to formulations of string theories where the duality symmetries
are manifest.

When such stringy geometries involve T-duality, they are called T-folds \cite{Hull:2004in}.
They have been studied mainly in the framework of
supergravity and Double Field They \cite{Hull:2009mi}. 
In order to go beyond and analyze their quantum aspects,
one may need the world-sheet approach based on conformal field theory (CFT).
The transitions  in the target space by T-duality
are represented on the world-sheet as
the twists by the T-duality transformations. 
Since they are generally left-right asymmetric,
the world-sheet theories fall into a particular class of asymmetric
orbifold CFTs. Such T-fold CFTs have been studied e.g.
in \cite{Flournoy:2005xe,HW,KawaiS1,KawaiS2,Anastasopoulos:2009kj,Bianchi:2012xz,
 Condeescu:2012sp,Condeescu:2013yma,SatohS,Tan:2015nja,Satoh:2015nlc,Sugawara:2016lpa}.

As is generally the case  for asymmetric orbifolds, the construction of the T-fold CFTs 
is not automatic. In addition to 
the modular invariance, there is an issue of the relative phases
of the action of the T-duality twist on the left- and right-movers \cite{HW,KawaiS1}, which may be 
regarded as an analog of the discrete torsions.  As an interesting consequence
of the explicit construction of a class of T-fold CFTs, 
it is  found that T-folds 
provide a simple setting to realize non-supersymmetric string vacua
with vanishing cosmological constant at least at one loop \cite{Satoh:2015nlc}. 
The mechanism there is extended
to a more general class of asymmetric orbifolds \cite{Sugawara:2016lpa}.
Through T-folds and related more general non-geometric backgrounds 
(monodrofolds \cite{Flournoy:2004vn}),
one can also explore the possibility that 
the world-sheet conformal interfaces \cite{Wong:1994np,Petkova:2000ip,Bachas:2001vj}, 
which may be regarded as fundamental from the world-sheet point of view,
can be applied to string theory \cite{SatohS}.
For the applications of the conformal interfaces to string theory, see e.g.  
\cite{Bachas:2007td,Satoh:2011an,Bachas:2012bj,Elitzur:2013ut}.

In spite of the developments on T-fold CFTs so far, 
we are still lacking their general construction, which is 
in contrast to the quite general analysis from the target space point of view by
supergravity.
The purpose of this paper is to advance a step in this direction and 
provide a systematic construction of the modular invariant partition functions 
of T-fold CFTs. A point of our construction is that we formulate the problem
based on the momentum lattices in order to control the modular properties 
of the partition functions in the twisted sectors from 
the asymmetric T-duality twist.
This allows us to consider the cases which are not reduced to
the free fermion construction as well.%
\footnote{
Based on the free fermion construction, 
systematic scans of of certain classes of T-fold CFTs have been
performed for type II \cite{Anastasopoulos:2009kj} and
heterotic \cite{Bianchi:2012xz} superstrings.
}

The condition that the T-duality twist acts in a single Hilbert space requires 
the background moduli of the torus compactification to be invariant 
under the T-duality transformation, i.e. at the fixed points,
as is found also in the supergravity analysis \cite{Dabholkar:2002sy,Dabholkar:2005ve}.
Imposing this condition, 
we first give the modular invariant partition functions for T-folds in bosonic string theory,
whose momentum lattices are associated with the Lie algebra lattices called 
Englert-Neveu lattices. The fixed-point condition  restricts the possible Lie algebras.
In the case where the T-duality acts as a simple chiral reflection in the right-mover,
we are left with the four cases, $A_1, D_{2r}, E_7, E_8$
among the simple simply-laced algebras.
Similar construction is also possible by using  Euclidean even self-dual lattices.
We then apply our construction to the T-folds in the $E_8 \times E_8$ heterotic string theory.
Including non-trivial phases in the T-duality twists,
the twisted partition functions in the originally intact left-mover are
represented by the building blocks which appeared in the bosonic-string case.
As simple examples,
we explicitly construct a class of the modular invariant partition functions of 
the T-fold CFTs parametrized by three integers.
The cases including the building blocks from $A_1$ and $E_7$
are not covered by the fermionization.

The rest of this paper is organized as follows: In section 2, we summarize the toroidal
compactification and T-duality in  bosonic string theory, which also serves as
fixing our notation and conventions. In section 3, we set up our problem and 
analyze the fixed-point (self-duality) condition of the T-duality transformations.
In section 4, we construct the modular invariant partition functions for T-folds
in bosonic string theory,
based on the Lie algebra lattices.
In section 5, we apply our construction to the T-folds in the heterotic string theory.
In section 6, we conclude with a summary and discussion.
In appendix A, we summarize the characters of the affine Lie algebras and our building blocks
for the modular invariant partition functions.

\section{Toroidal compactification and T-duality}

Let us consider the bosonic string theory where $d$ coordinates are compactified
on a $d$-dimensional torus $T^d$. We basically follow the conventions 
in \cite{Giveon:1994fu,Kugo:1992md}: The compactified 
coordinates $X^i$ $(i = 1, ..., d)$ have a periodicity $X^i \approx X^i + 2\pi$. 
The constant background fields, 
the metric $G_{ij}$ and the anti-symmetric tensor $B_{ij}$,  are organized into a matrix, 
\be
\label{EGB}
  E_{ij} := G_{ij} + B_{ij} \comma
\ee
which forms the background moduli of the compactification.
The metric and anti-symmetric tensor are its symmetric and anti-symmetric part, 
respectively, i.e. $G = (E+E^t)/2$, $B=(E-E^t)/2$, where we have suppressed indices. 
The vielbeins are defined so that
\be
 \label{eeG}
    \sum_{a,b=1}^d \delta_{ab} e^a_i e^b_j = 2 G_{ij} \comma \quad
   \sum_{a=1}^d e^a_i e^{*j}_a = \delta^i_j \comma 
\ee
implying $\sum_{a,b=1}^d \delta^{ab} e^{*i}_a e^{*j}_b = \frac{1}{2} G^{ij}$.
The space-time indices $i,j$ are converted to those of the tangent
space $a,b$ by $ e^a_i/\sqrt{2},  \sqrt{2} e^{*i}_a$, and 
they are  lowered and raised by $G_{ij}, G^{ij}$ and $\delta_{ab}, \delta^{ab}$.

The world-sheet Hamiltonian takes the form,
\be
  H= L_{0L}+ L_{0R} \comma
\ee
where 
\be
L_{0L}  := \frac{1}{2} p_{L}^2 + N
      \comma \quad  
    L_{0R} := \frac{1}{2} p_{R}^2 + \tilde{N} \comma \nn
\ee
\be
\label{pRL}
  p_{La}(E) :=  e^{*i}_a [ n_i - E_{ij} w^j] 
   \comma \quad 
   p_{Ra}(E) :=  e^{*i}_a [n_i + E^t_{ij} w^j ] 
     \comma \
\ee
with $p_{L/R}^2 = p_{L/Ra} \delta^{ab}  p_{L/Rb} $ and $n_i, w^j \in \bbZ$ 
being the momentum and winding numbers.
The dependence on the background $E_{ij}$
has been  indicated explicitly in the momenta $p_{L/R}$. 
The transpose of $E_{ij}$ reads $E^t_{ij} = G_{ij} - B_{ij}$.
The remaining terms $N, \tilde{N}$ are the number operators for the oscillator modes,
\be
 N(E) = \sum_{m >0} \alpha_{-m}^i(E)G_{ij} \alpha_m^{j}(E)
 \comma \quad 
 \tilde{N}(E)= \sum_{m >0} \tilal_{-m}^i(E)G_{ij} \tilal_m^{j}(E) \comma
\ee
which take the values of non-negative integers. 
The partition function then takes the form
\be
\label{twistZ}
  \Tr  \Bigl[ \, q^{L_{0L}-\frac{d}{24}} 
   \qbar^{L_{0R}-\frac{d}{24}}\, \Bigr]
   = \frac{1}{|\eta(\tau)|^{2d}} \sum_{p_L, p_R} q^{\frac{1}{2}p_L^2} \bar{q}^{\frac{1}{2}p_R^2}
   \comma
\ee
where 
$q=e^{2\pi i \tau}$ and  $\tau = \tau_1 + i \tau_2$ ($\tau_1 \in \bbR, \tau_2 >0$) is the modulus of
the torus.
The sum over the zero-modes are regarded as that over the Lorentzian lattice $\Lambda$
which is formed by the pair of the momenta $(p_L, p_R)$, and equipped with 
the Lorentzian inner product, $(p_L, p_R) \circ (p'_L, p'_R) = p_L p'_L - p_R p'_R$.
Since this lattice is even self-dual, i.e. $(p_L, p_R) ^2 \in 2 \bbZ$ and $\Lambda = \Lambda^*$
(dual lattice), the above partition function is modular invariant.
The zero-mode part of $H$ is concisely expressed as
\be
\label{H0}
  H_0 = \frac{1}{2} (p_L^2 + p_R^2) = \frac{1}{2} v^t M(E) v \comma
  \quad 
\ee
where $v := (w^i, n_j)^t$ and 
\be
\label{MEZ}
 M(E) := \left(\begin{array}{cc} G - BG^{-1}B & BG^{-1} \\ -G^{-1} B & G^{-1} \end{array}\right) 
 \period
\ee
This Hamiltonian keeps its form under the canonical map 
\cite{Kugo:1992md} associated with an $O(d,d,\bbZ)$
matrix. 
To see this, we first note that 
an $O(d,d)$ matrix $g$ is defined as a $2d \times 2d $ matrix satisfying 
\be
  \label{defOdd}
  g^t J g = J \comma \quad J := \left(\begin{array}{cc}0 & I_{d} \\ I_{d} & 0\end{array}\right)
  \comma
\ee
with $I_k$ being the $k$-dimensional identity matrix.  One finds that
$M(E) \in O(d,d,\bbR)$.
We then consider  two backgrounds $E$ and $E'$ which are related 
to each other by the  $O(d,d,\bbZ)$ transformation,
\be\label{EE'}
  E \to E' = g(E) := (a E + b)(cE+d)^{-1} \comma
  \ee
for $g \in O(d,d,\bbZ)$ of the form,
\be
\label{gabcd}
  g = \left(\begin{array}{cc} a & b \\c & d\end{array}\right) \period
\ee
This in turn implies the map
of the metric,  
\be
  G \to G' = \gamma_L^\star G \gamma_L^{-1}  =  \gamma_R^\star G \gamma_R^{-1}  \comma
\ee
and that for the vielbein, 
\be
\label{etrans}
   {e}_i^a \to {e'}_i^a =  e_j^a (\gamma_L^{-1})_{ji} = e_j^a (\gamma_R^{-1})_{ji} \comma
   \ee
up to orthogonal transformations in the tangent space. Here, we have 
defined 
\be
  \gamma_L(E):= (d-cE^t) \comma \qquad  \gamma_R(E) := (d+cE)  \comma
\ee
and $O^\star := (O^t)^{-1}$ for any invertible matrices, and used $(E')^t = (a E^t - b)(-cE^t+d)$.
The canonical map acts 
on the oscillators as
\be
\label{almap}
   \alpha_m(E) \to \gamma_L^{-1}(E) \alpha_m(E') \comma \quad
   \tilal_m(E) \to \gamma_R^{-1}(E) \tilal_m(E') \period
\ee
These are valid also for the zero-modes with the level $m=0$, 
\be
   \alpha_0(E) = \frac{1}{\sqrt{2}} G^{-1} (n-Ew)
   \comma \quad 
    \tilal_0(E) = \frac{1}{\sqrt{2}} G^{-1} (n+E^tw)  \comma
\ee
and thus  (\ref{almap}) is translated into 
\be
  \label{hatpTrans}
   \hatp_{L}(E) \to   \gamma_L^t(E)  \, \hatp_{L}(E') \comma \quad
   \hatp_{R} (E)\to \gamma_R^t(E)  \, \hatp_{R}(E') \comma  
\ee
in terms of 
\be
\label{hatp}
   \hatp_{Lj}(E):= e_i^a p_{La}(E) = (n-wE^t)_j \comma \quad 
  \hatp_{Rj}(E):= e_i^a p_{Ra}(E) = (n+ wE)_j \period
\ee
By these transformation rules, 
the number operators are mapped as
\be
  N(E) \to N(E') \comma \quad \tilde{N}(E) \to \tilde{N}(E') \comma
\ee
whereas
\be
\label{H0Trans}
  H_0(E) \to \frac{1}{2} v^t gM(E) g^t v =H_0(E') \comma
\ee
where we have used $M(E') = gM(E) g^t$. 
These indeed show that the form of $H(E)$ is kept intact.
The transformation (\ref{H0Trans}) is regarded as either that of $M(E)$ with $v$ kept fixed
or that of $v$ with $E$ kept fixed.
Since $g \in O(d,d,\bbZ)$ and hence the integer-valued vector $g^t v$ can be renamed as $v$,
one confirms that the spectrum is invariant under the map.

\section{Partition functions for T-folds} 
\label{sec.PfnTf}

We now consider the asymmetric orbifolds by the $O(d,d,\bbZ)$ T-duality transformations 
discussed in the previous section.
In particular, we start from the target space,
\be
 M \times \bbR \times T^d \comma
\ee
and twist the strings on it by the operator
\be
\label{twist}
 \sigma = \calT_{2\pi R} \otimes g  \period 
\ee 
Here $ \calT_{2\pi R}$ stands for the shift by $2\pi R$ in  $ \bbR$,
and $g  \in O(d,d,\bbZ)$ for the T-duality twist acting on $T^d$. $M$ is other non-compact part.
Consequently, we are considering 
a class of non-geometric backgrounds, i.e. T-folds, 
where $\bbR$ twisted by $\calT_{2\pi R}$ provides  the `base' circle $S^1_R$
with radius $R$ , while $T^d$ is its `fiber'.

\subsection{World-sheet partition functions}
\label{subsec.WsPf}

In order to construct the world-sheet torus partition functions describing the strings on 
the above T-folds,
we start with the partition function for the $\bbR \times T^d$ part with the $m$-fold temporal twist,
\be
\label{Psum0m}
   Z_{(0,m)}(\tau)
   := \Tr  \left[\sigma^m \, q^{L_0-\frac{c}{24}} 
   \qbar^{\tilL_0-\frac{c}{24}}\, \right] \period
\ee
Here, $L_0, \tilL_0$ and $c$ are the Virasoro generators and the central charge, respectively, 
and the trace is taken over the untwisted Hilbert space.
The trace in the base 
part is evaluated as
\be
\Tr_{\rm base} \left[\left(\calT_{2\pi R}\right)^m \, q^{L_{0}^{\rm base}-\frac{1}{24}} 
\qbar^{\tilL_{0}^{\rm base}-\frac{1}{24}}\, \right]
= Z_{R, (0,m)}(\tau) \comma
\ee
where 
\be
\label{S1Psum}
  Z_{R, (w,m)}(\tau) = 
\frac{R}{\sqrt{\tau_2}|\eta(\tau)|^2} 
e^{-\frac{\pi R^2}{\tau_2}|w\tau+m|^2} 
\ee
is the partition function for a free boson on $S^1_R$  in the  winding 
sector with 
the spatial and temporal winding number $w, m \in \bbZ $, respectively.
 $\eta(\tau)$ is the Dedekind $\eta$ function.
If the twist acted on the $\bbR$
and the $T^d$ part independently,
the partition function in the base part would be $\sum_{w,m \in \bbR} Z_{R, (w,m)}$, 
giving the ordinary partition function for a compactified free boson.
Denoting the fiber part as $Z^{T^d}_{(0,m)}$, 
the trace  in (\ref{Psum0m}) is written
as $ Z_{(0,m)} = Z_{R, (0,m)} Z^{T^d}_{(0,m)} $.

The partition functions in the base part 
transform covariantly under the modular transformations,
\begin{align}
& \left. Z_{R, (w,m)}(\tau)\right|_T \left(\equiv Z_{R, (w,m)}(\tau+1)\right) =  Z_{R, (w,w+m)}(\tau) \comma
\nn
& \left. Z_{R, (w,m)}(\tau)\right|_S \left(\equiv  Z_{R, (w,m)}(-1/\tau)\right) = Z_{R, (m,-w)}(\tau) \period
\label{ModCov}
\end{align}
These form a particular representation of the modular group.
If the fiber part $Z^{T^d}_{(0,m)}$ satisfies the same form of the modular
covariance,
\be
\label{ModCov2}
\left. Z_{(w,m)}^{T^d}(\tau)\right|_T 
=  Z_{(w,w+m)}^{T^d}(\tau) \comma  \qquad
\left. Z_{(w,m)}^{T^d}(\tau)\right|_S
=  Z_{(m,-w)}^{T^d}(\tau) \comma
\ee
they give $Z^{T^d}_{(w,m)}$ with general winding numbers.
Summing up all, 
the total partition function in such a case,
\be
\label{Ztot}
   Z(\tau)  
     = Z_{M} (\tau) \sum_{w,m \in \bbZ}  Z_{R, (w,m)}(\tau)\,
Z^{T^d}_{(w,m)}(\tau) \comma
\ee
becomes modular invariant. Here, 
the first factor  $Z_{M}$ is the contribution from $M$ in the background, which is 
assumed to be modular invariant itself. 

In this argument, a non-trivial step 
for constructing the modular invariant $Z(\tau)$
is to find the fiber partition functions with the desired covariance (\ref{ModCov2}).
We see that a formulation based on the momentum lattices is useful
to control the modular properties of the fiber part for that purpose.

\subsection{Fixed points of T-duality transformations}

In the fiber part, the twist operator $\sigma$ acts as a T-duality transformation.
In general, T-duality connects different (but equivalent) world-sheet theories, and thus 
in order for the twist to be well-defined in a single Hilbert space, it has to 
be self-dual.%
\footnote{In this paper, we use  `self-dual' to express both the 
`self-dual' T-duality transformation satisfying (\ref{ESD}) and the `self-dual' lattice
satisfying $\Lambda = \Lambda^*$ as below (\ref{twistZ}). 
}
In other words, 
the CFTs for T-folds are defined at the fixed points 
of the moduli space under the T-duality transformations. 
This also conforms to the supergravity analysis
\cite{Dabholkar:2002sy,Dabholkar:2005ve}.
Given the transformation rule (\ref{EE'}), 
this condition is represented as
\be
\label{ESD}
   E = (aE+b)(cE+d)^{-1} \comma 
\ee
for $g$ of the form (\ref{gabcd}). This also implies  the invariance of the metric,
\be
\label{Ginv}
  G = \gamma_{L/R}^t G \gamma_{L/R} \period 
\ee 
Denoting the momentum squared as
\be
  p_L^2(E) =  \hatp^t_L(E)  \frac{1}{2}G^{-1}  \hatp_L(E) \comma \quad
   p_R^2(E) = \hatp^t_R(E)  \frac{1}{2}G^{-1}  \hatp_R(E) \comma
\ee
one finds that $p_{L/R}^2$ is separately invariant in the self-dual case.

To read off the form of the $O(d,d,\bbZ)$ element implementing the self-dual
transformation, we rewrite the transformation 
(\ref{hatpTrans}) in the form,
\be
 \label{hatpTrans2}
  \begin{pmatrix}
     \hatp_{Lj}(E)    \\
     \hatp_{Rj}(E)   
\end{pmatrix}
= P(E) v \to \Gamma(E)^t P(E) v =: P(E) g_{SD}^t v \comma
\ee
where $v = (w^i, n_j)^t$ as before, 
$P(E):= \begin{pmatrix}
     -E &   I_d \\
      E^t &  I_d
\end{pmatrix} $,
and 
$ \Gamma(E) := 
\begin{pmatrix}
     \gamma_L &   0 \\
      0 &  \gamma_R
\end{pmatrix}
$.
After the map in the above, one has $\hatp_{L/R}(E)$ instead of $\hatp_{L/R}(E')$  due 
to the self-duality. Comparing this to the map of the Hamiltonian (\ref{H0Trans}), 
one finds that $g_{SD}$ in the above represents  
the corresponding $O(d,d)$ element. 
Its explicit form is \cite{Erler:1996zs,Tan:2015nja}
\eqb
\label{gSD}
  g_{SD} \Eqn{=} P^{t} \Gamma P^\star \nonumber \\
  \Eqn{=}  \frac{1}{2} \begin{pmatrix} 
      \gamma_L^{\star} + \gamma_R^{\star} - B \gamma_-G^{-1}
    & -(\gamma_L^{\star} + \gamma_R^{\star} - B \gamma_- G^{-1})B  - G\gamma_- + B\gamma_+ \\ 
   - \gamma_- G^{-1}
    & \gamma_- G^{-1}B + \gamma_+ 
    \end{pmatrix} \comma
\eqe
with $\gamma_\pm := \gamma_L \pm \gamma_R$. Here,
we have used 
$ P(E)^{-1} = \frac{1}{2}
\begin{pmatrix}
      -I_d & I_d   \\
      E^t& E 
\end{pmatrix}
\begin{pmatrix}
      G^{-1} & 0   \\
       0 & G^{-1}
\end{pmatrix}
$, and the invariance of $G$ (\ref{Ginv}).
This gives a necessary condition on the form of the self-dual transformation.
Using the invariance of $G$, one can check that $g_{SD} \in O(d,d,\bbR)$.
Thus, if its components are  integer-valued, $g_{SD}$ provides a proper $O(d,d,\bbZ)$ self-dual
transformation.

\subsection{Fiber twist}
\label{subsec.Ftw}

As a simple example of (\ref{gSD}), we consider in this paper the case where 
\be
\label{gammaI}
    \gamma_L = I_d \comma \qquad \gamma_R =  -I_d
    \comma
\ee
and hence
\be
   g_{SD} = -\left(\begin{array}{cc} BG^{-1} & -BG^{-1}B+G \\ G^{-1}& -G^{-1} B 
      \end{array}\right) 
      =  -M(E) J \period
\ee
This is a $\bbZ_2$ element,  $g_{SD}^2 = I_{2d}$, since
$g^{-1} = J g^t J$ for $g \in O(d,d)$ and $M(E)^t  = M(E)$.
One can explicitly check that it induces a self-dual transformation $E\to E'=E$.
A sufficient condition for the integer-valuedness  of $g_{SD}$ is
$E_{ij}, G^{ij}/2 \in \bbZ$, which follows from the  product form in (\ref{gSD}).
When $E_{ij}$ is triangular e.g. $B_{ij} = G_{ij}$ $(i >j)$, it is also 
sufficient that $E_{ij}, G^{ij} \in \bbZ$. This is confirmed by noting that
$E_{ij} \in \bbZ$ implies  $2G_{ij}, G_{ii} $ (no sum) $\in \bbZ$ 
and that the products of $B_{ij}$ and matrices 
are rewritten as 
\begin{equation}
\sum_j B_{ij} M_{jk} = \sum_j G_{ij} M_{jk} - 2 \sum_{i<j} G_{ij} M_{jk} - G_{ii} M_{ik},
\end{equation}
which is used for $M = G^{-1}$ or $G^{-1}B$.
We note that in general the above $g_{SD}$ does not correspond 
to the $G \leftrightarrow G^{-1}$ ($R \leftrightarrow 1/R$) 
duality, in spite of the forms of $\gamma_{L}, \gamma_{R}$.

In this case, from the transformation (\ref{almap}), 
i.e. $(\alpha_m, \tilal_m) \to (\alpha_m, -\tilal_m)$,
it follows that  the oscillator contribution to the twisted partition function in the fiber part
becomes $\bigl[ \overline{2\eta(\tau)/\theta_2(\tau)} \bigr]^{d/2} \times \eta(\tau)^{-d}$. 
In the zero-mode part,
the right momenta are projected out, $\hatp_R =0$, which implies that $ n= -E^t w$ and hence
we are left with the Euclidean lattice sum in the left-moving sector with this constraint. 
Taking into account 
$g_{SD}^m = 1$ $(m \in 2 \bbZ)$
and $g_{SD}^m=g$ $(m \in 2 \bbZ+1)$ in the untwisted sector ($w=0$), we have
\be
 \label{FiberTw}
   Z^{T^d}_{(0,m)}(\tau) 
   = \Tr  \left[ g_{SD}^m \, q^{L_{L0}-\frac{d}{24}} 
   \qbar^{L_{R0}-\frac{d}{24}}\, \right]
   = 
   \overline{\TTheta_{34}}^{d/2}(\tau) \cdot 
   \frac{1}{\eta(\tau)^d}
     \sum_{wE \in \bbZ^{d}} q^{w^t G w} \comma
\ee
for $(m \in 2 \bbZ+1)$, 
where we have used $\theta_2\theta_3\theta_4 = 2 \eta^3$ and defined 
\be
\label{thetapq}
  \TTheta_{pq}(\tau) := \frac{\theta_p(\tau) \theta_q(\tau)}{\eta(\tau)^2} \period
\ee
Below, we show that further choosing appropriate backgrounds yields 
the partition functions with the desired modular covariance (\ref{ModCov2}), 
and thus the modular invariant total partition functions.

\section{T-folds from lattices}
\label{sec.TLat}

In this section, we show that 
one can systematically construct the fiber partition functions with the desired 
modular covariance
by choosing the background moduli  $E_{ij}$ associated with the Lie algebra lattices, 
namely, sublattices of the weight lattice of a semi-simple Lie algebra.
We first discuss the case of Englert-Neveu lattices \cite{Englert:1985ws} 
for simply-laced Lie algebras
and then the case of  Euclidean even self-dual lattices,
both of which are straightforwardly realized by the momentum lattices of bosonic strings.
For a review on the lattices in relation to string theory, see for example \cite{Lerche:1988np}.

\subsection{Lie algebra lattices and Englert-Neveu lattices}
\label{subsec.LAL}

We consider the background with an affine symmetry of level one $\widehat{X}_1$ 
for  a semi-simple
simply-laced Lie algebra $X$ which is realized by \cite{Elitzur:1986ye,Giveon:1994fu}
\be
  \label{EC}
  E_{ij} = C_{ij} \ (i>j) \comma \qquad 
   E_{ii} =  \frac{1}{2} C_{ii}  \comma \qquad  E_{ij} = 0 \ (i< j) \period
\ee 
Here, $C_{ij}$  is the Cartan matrix of $X$, 
and the indices are not summed in the middle equation. 
The simple roots are normalized so that their norms are equal to two.
In this background, $ e_i \cdot e_j =  2 G_{ij} = C_{ij}$ (for any $i,j$),
and thus $e_i$ are the simple roots, whereas the duals  $e^{*i}$ are the fundamental weights.
Since $E_{ij} \in \bbZ$, the sum over the momenta in (\ref{pRL}) becomes that over
the weight lattice. 
Furthermore, since $p_{Ra} - p_{La} = e_j^a w^j$, the weights in the left- and
the right-movers belong to the same conjugacy class. 
Up to this constraint, 
one can confirm by using the inverse of $P(E)$ in (\ref{hatpTrans2}) that 
the summation reduces to the independent ones in each of the left- and right-movers.
This gives an explicit realization of the Lorentzian even self-dual (Narain) lattice
$(p_L,p_R)$ of the type called the Englert-Neveu lattice \cite{Englert:1985ws}.

Thus, without twists, the relevant partition function is given by
the sum of the diagonal combinations of 
of the level-one affine Lie algebra characters for $X$,
\be
   \sum_\alpha \, \bigl\vert \chi_\alpha^{X} (\tau) \bigr\vert^2 \comma
\ee
where 
\be
  \chi_\alpha^{X}(\tau) 
  := \frac{1}{\eta(\tau)^r} \sum_{\blam_\alpha \in \Lambda_{(\alpha)}^X} 
  q^{\frac{1}{2} \blam_\alpha^2} \comma
\ee
and $r$ is the rank of $X$.
The summation is taken over the weights $\blam_\alpha$ belonging to a conjugacy class 
$\Lambda_{(\alpha)}^X$, i.e. an element of the coset $\Lambda^*_X/\Lambda_X$
 labeled by $\alpha$, where 
 $\Lambda^*_X$ and $\Lambda_X$ are  the weight and the root lattice
of $X$, respectively.
A conjugacy class $\Lambda_{(\alpha)}^X$ also corresponds to 
an integrable representation of the affine Lie algebras at level one $\widehat{X}_1$.

For our purpose,  a useful fact on the Lie algebra lattices is that 
these characters form a finite dimensional representation of the modular group,
which is summarized as 
\be
\label{chMT}
\left. \chi_\alpha^{X}(\tau)\right|_T 
= T^{X}_{\alpha\beta} \, \chi^{X}_\beta(\tau) \comma \qquad
\left. \chi_\alpha^{X}(\tau)\right|_S
= S^{X}_{\alpha\beta} \, \chi^{X}_\beta(\tau) \period
\ee
The modular matrices here are given by
\be
   T^{X}_{\alpha\beta} = e^{-\pi i (\frac{r}{12} - \blam^2_\beta)} \delta_{\alpha\beta} \comma
   \qquad
   S^{X}_{\alpha\beta} = \frac{1}{\sqrt{N_c}}  e^{2\pi i \blam_\alpha\cdot \blam_\beta }  \comma
\ee
in terms of the weight vectors $\blam_\alpha \in \Lambda_{(\alpha)}^X$ 
and the number of the conjugacy classes $N_c$. 

Now let us return to the construction of the partition functions for T-folds.
First, we note that,
since $E_{ij} \in \bbZ$, the constraint  $wE \in \bbZ^d $ in (\ref{FiberTw}) is automatically
satisfied, and hence the summation becomes that over  the root lattice.
This enables us to utilize the above modular covariance to derive the partition
functions in the twisted sectors.

Next, we note that the condition discussed in the previous section
 that $g_{SD}$ in (\ref{gSD}) with $\gamma_{L/R}$ in (\ref{gammaI}) 
be integer-valued constrains the possible Englert-Neveu lattices.
In particular,  due to the condition that 
$G^{-1} = 2 C^{-1}$ is integer-valued, 
we are left with 
\be
 \label{SDG}
  A_1 \comma \quad D_r \ (r: {\rm even}) \comma \quad E_7 \comma \quad E_8 \comma
\ee
among the simple simply-laced Lie algebras.
Since $E_{ij} $ is triangular and its elements are integral, $ E_{ij} \in \bbZ$, 
one finds that 
 $g_{SD}$ is indeed integer-valued for the algebras
in (\ref{SDG}) and for their products, as discussed in section \ref{subsec.Ftw}.
Since any background realized by (\ref{EC}) is a fixed point under some non-trivial
$O(d,d,\bbZ)$ transformation \cite{Giveon:1988tt,Giveon:1994fu}, one may consider other 
simply-laced Lie algebras. 
It may also be possible to consider the $\bbZ_N$ elements of $O(d,d,\bbZ)$ as in \cite{Tan:2015nja}.
However, the corresponding twists are more involved than
(\ref{gammaI}).

By starting from the twisted partition functions $Z_{(0,m)}^{T^d}$ given in (\ref{FiberTw}) and using  
the modular properties (\ref{chMT}), 
we can now 
uniquely determine the  whole building blocks $Z_{(w,m)}^{T^d}$ for the cases  \eqn{SDG}
including the suitable phase factors to achieve the modular covariance (\ref{ModCov2}).
We concisely call this prescription and the resultant blocks as the {\em `modular completions'} \footnote
   {In this paper we use this terminology in the sense different from e.g.  \cite{Eguchi:2010cb}. } 
in the arguments below.
Combining these $Z_{(w,m)}^{T^d}$ with other parts, we obtain the modular invariant
partition functions of the form (\ref{Ztot}) for the T-fold CFTs.
One can also utilize products 
$ Z_{(w,m)}^{T^{d+d'}} = Z_{(w,m)}^{T^d} \times Z_{(w,m)}^{T^{d'}}$
where each factor corresponds to any of the algebras in (\ref{SDG}).

We list the result of $Z_{(w,m)}^{T^d}$ in each case below.
The corresponding Lie algebras are explicitly denoted there.
Among the list, the cases for $D_2$ and $D_4$ appeared e.g.
in \cite{Satoh:2015nlc,Sugawara:2016lpa}. 
The appearance of  $A_1$ and $E_7$ may also be of interest, since such a case
is not  covered by the ordinary fermionization. 
We note that  the action of $g_{SD}$ in the partition functions below 
is $\bbZ_2$ in the untwisted  Hilbert space with $a=0$, 
which is in accord with the supergravity picture.
In  the twisted Hilbert spaces with $a \neq 0$, this is however not the case,
except for $D_r$ $(r \in 8 \bbZ)$ and $E_8$.
A related discussion on the modular covariance in the $A_1$ case is 
found in \cite{Aoki:2004sm}.
In the following, we denote the fiber torus corresponding to $X$ by
$T^d[X]$.

\paragraph{Partition functions for  $A_1$ :}
There are two conjugacy classes $\Lambda_{(\alpha)}^{A_1}$ 
with $\alpha = 0,1$, which include the spin $\alpha/2$-representation. 
The norms of the weight $\blam_\alpha$ for these conjugacy classes are
$\blam_0^2 = 0$, $\blam_1^2 = 1/2$ (mod $2$) and $\blam_0 \cdot \blam_1 = 0$ 
 (mod $1$). Thus, 
\be
  T^{A_1}_{\alpha\beta} = e^{-\frac{\pi}{12}i} \diag(1,i)
            \comma \qquad
  S^{A_1}_{\alpha\beta} = \frac{1}{\sqrt{2}} \begin{pmatrix}
                                   1 & 1 \\
                                   1 & -1
                                  \end{pmatrix} 
    \period
\ee
Since $ Z^{T^1[A_1]}_{(0,m)}(\tau)  = \overline{\TTheta_{34}(\tau)}^{1/2} \chi^{A_1}_0(\tau)$
$(m \in 2 \bbZ +1)$, the modular completion yields
\be
\label{A1cov}
    Z^{T^1[A_1]}_{(a,b)}(\tau) 
    = \left\{ 
            \begin{array}{ll}
             \bigl{|}    \chi^{A_1}_0(\tau)  \bigr{|}^2 +   \bigl{|}  \chi^{A_1}_1 (\tau)  \bigr{|}^2
       &
       (a \in 2 \bbZ \comma b \in 2\bbZ)  \comma \\
            e^{\frac{\pi i}{8}ab^3} \, \overline{ \TTheta_{34}(\tau)}^{1 \o 2} 
         { 
         \cdot \frac{1}{2} } \Bigl[  \chi^{A_1}_+ (\tau) +  i^a \chi^{A_1}_- (\tau) \Bigr]
       &
       (a \in 2 \bbZ \comma b \in 2\bbZ+1)  \comma \\
         e^{-\frac{\pi i}{8}a^3b} \, \overline{ \TTheta_{23}(\tau)}^{1 \o 2} 
         \cdot
          { 
          \frac{1}{\sqrt{2}} }
          \Bigl[ \chi^{A_1}_0(\tau) +  i^b \chi^{A_1}_1(\tau)  \Bigr] 
       &
       (a \in 2 \bbZ+1 \comma b \in 2\bbZ)  \comma \\
         e^{-\frac{\pi i}{8}a^3b} \,  \overline{ \TTheta_{24}(\tau)}^{1 \o 2} 
         \cdot
       { 
          \frac{1}{\sqrt{2}} }
         \Bigl[  \chi^{A_1}_0(\tau) +  i^{a+b-1} \chi^{A_1}_1(\tau)  \Bigr] 
       &
       (a \in 2 \bbZ+1 \comma b \in 2\bbZ+1) \comma 
            \end{array}
    \right.
 \ee
which indeed satisfies the desired modular covariance (\ref{ModCov2}).
Here, $\chi^{A_1}_\pm := \chi^{A_1}_0 \pm \chi^{A_1}_1 $, 
and the explicit forms of the characters $\chi^{A_1}_{0,1}$ are 
found in (\ref{A1 ch}) in the appendix.

\paragraph{Partition functions for $D_r  \ (r: {\rm even})$ :}
There are four conjugacy classes $\Lambda_{(\alpha)}^{D_r}$,  which include  
 the vacuum, vector, spinor or conjugate-spinor representation.
We label these by $\alpha = 0, v,s,c$, respectively.
A representative weight in each conjugacy class is
$(0,...,0)$, $(1, 0, ...., 0)$, $(\pm 1/2, \pm 1/2, ...)$ with even or odd 
number of minus signs. Thus,
\be
  T^{D_r}_{\alpha\beta} =  e^{-\frac{\pi i}{12}r} \diag(1,-1, e^{\frac{\pi i}{4} r}, e^{\frac{\pi i}{4} r})
   \comma \qquad
      S^{D_r}_{\alpha\beta} = \frac{1}{2} \begin{pmatrix}
                                   1 & 1 & 1 & 1 \\
                                   1 &  1 &-1 & -1 \\
                                    1 & -1 &   i^r & -i^r \\
                                    1 & -1 &  -i^r & i^r
                                  \end{pmatrix} 
\period
\ee
Since $ Z^{T^r[D_r]}_{(0,m)}(\tau)  = \overline{\TTheta_{34}(\tau)}^{r/2} \chi^{D_r}_0(\tau)$
$(m \in 2 \bbZ +1)$, the modular completion yields
\be
    Z^{T^r[D_r]}_{(a,b)}(\tau) = \left\{
     \begin{array}{ll}
     \displaystyle{  \frac{1}{2|\eta(\tau)|^{2r}} 
            \left[ \big| \theta_3(\tau) \big|^{2r} +  \big|\theta_4(\tau) \big|^{2r} 
            +  \big|\theta_2(\tau) \big|^{2r} \right]
           }
          &
       (a \in 2 \bbZ \comma b \in 2\bbZ)  \comma \\
         e^{\frac{\pi i r}{8}ab} \, \overline{ \TTheta_{34}(\tau)}^{r \o 2} 
         {\displaystyle \cdot \frac{1}{2\eta(\tau)^r} \left[  \theta_3(\tau)^r 
       +  e^{\frac{\pi i r}{4}a} \theta_4(\tau)^r \right]}
       &
       (a \in 2 \bbZ \comma b \in 2\bbZ+1)  \comma \\
         e^{-\frac{\pi i r}{8}ab} \, \overline{ \TTheta_{23}(\tau)}^{r \o 2} 
          {\displaystyle \cdot \frac{1}{2\eta(\tau)^r} \left[  \theta_3(\tau)^r
       +  e^{\frac{\pi i r}{4}b} \theta_2(\tau)^r \right]}
       &
       (a \in 2 \bbZ+1 \comma b \in 2\bbZ)  \comma \\
         e^{-\frac{\pi i r}{8}ab} \,  \overline{ \TTheta_{24}(\tau)}^{r \o 2} 
         {\displaystyle \cdot \frac{1}{2\eta(\tau)^r} \left[  \theta_4(\tau)^r 
       +  e^{\frac{\pi i r}{4}(a+b-1)} \theta_2(\tau)^r \right]}
       &
       (a \in 2 \bbZ+1 \comma b \in 2\bbZ+1) \comma
   \end{array}
   \right.
\label{Drcov}
\ee
satisfying the modular covariance (\ref{ModCov2}).
The theta functions are related to the $D_r$ characters as in (\ref{Dr ch}).

\paragraph{Partition functions for $E_7$ :}
There are two conjugacy classes $\Lambda_{(\alpha)}^{E_7}$, 
which include the vacuum or the  ${\bf 56}$ representation.
We label these by $\alpha=0$ and $\alpha=1$, respectively. 
The norms of the corresponding weights are
$\blam_0^2 = 0$, $\blam_1^2 = 3/2$ (mod $2$) and $\blam_0 \cdot \blam_1 = 0 $ 
 (mod $1$). Thus, 
\be
  T^{E_7}_{\alpha\beta} = e^{-\frac{7\pi}{12}i} \diag(1,-i)
   \comma \qquad
      S^{E_7}_{\alpha\beta} = \frac{1}{\sqrt{2}} \begin{pmatrix}
                                   1 & 1 \\
                                   1 & -1
                                  \end{pmatrix} 
    \period
\ee
Since $ Z^{T^7[E_7]}_{(0,m)}(\tau)  = \overline{\TTheta_{34}(\tau)}^{7/2} \chi^{E_7}_0(\tau)$
$(m \in 2 \bbZ +1)$, the modular completion yields
\be
    Z^{T^7[E_7]}_{(a,b)}(\tau) 
    = \left\{ 
            \begin{array}{ll}
            \bigl{|}    \chi^{E_7}_0(\tau)  \bigr{|}^2 +   \bigl{|}  \chi^{E_7}_1 (\tau)  \bigr{|}^2
       &
       (a \in 2 \bbZ \comma b \in 2\bbZ)  \comma \\
            e^{\frac{7\pi i}{8}ab^3} \, \overline{ \TTheta_{34}(\tau)}^{7 \o 2} 
         {
         \cdot \frac{1}{2} } \Bigl[  \chi^{E_7}_+ (\tau) +  (-i)^{a} \chi^{E_7}_- (\tau) \Bigr]
       &
       (a \in 2 \bbZ \comma b \in 2\bbZ+1)  \comma \\
         e^{-\frac{7\pi i}{8}a^3b} \, \overline{ \TTheta_{23}(\tau)}^{7 \o 2} 
          {
         \cdot \frac{1}{\sqrt{2}} } 
          \Bigl[ \chi^{E_7}_0(\tau) +  (-i)^b \chi^{E_7}_1(\tau)  \Bigr] 
       &
       (a \in 2 \bbZ+1 \comma b \in 2\bbZ)  \comma \\
         e^{-\frac{7\pi i}{8}a^3b} \,  \overline{ \TTheta_{24}(\tau)}^{7 \o 2} 
         {
         \cdot \frac{1}{\sqrt{2}} }  
         \Bigl[  \chi^{E_7}_0(\tau) +  (-i)^{a+b-1} \chi^{E_7}_1(\tau)  \Bigr] 
       &
       (a \in 2 \bbZ+1 \comma b \in 2\bbZ+1) \comma 
            \end{array}
    \right.
\label{E7cov}
\ee
satisfying the modular covariance (\ref{ModCov2}).
Here , 
$\chi^{E_7}_\pm := \chi^{E_7}_0 \pm \chi^{E_7}_1 $, 
and the  explicit forms of the characters are found in (\ref{E7 ch 0}) and (\ref{E7 ch 1}).

\paragraph{Partition functions for $E_8$ :}
There is only one conjugacy class including the vacuum representation,
which we label by $\alpha=0$. 
The norms of the weights is
$\blam_0^2 = 0$ (mod $2$). This is an even self-dual lattice and hence
the modular property is trivial up to the phases coming from the eta functions for 
the oscillator part,
\be
  T^{E_8}_{\alpha\beta} = e^{-\frac{8\pi}{12}i} 
   \comma \qquad
      S^{E_8}_{\alpha\beta} =  1  
    \period
\ee
Since $ Z^{T^8[E_8]}_{(0,m)}(\tau)  = \overline{\TTheta_{34}(\tau)}^4 \chi^{E_8}_0(\tau)$
$(m \in 2 \bbZ +1)$, the modular completion yields
\be
    Z^{T^8[E_8]}_{(a,b)}(\tau) = 
    \left\{
     \begin{array}{ll}
          \bigl{|}  \chi^{E_8}_0(\tau) \bigr{|}^2
       &
       (a \in 2 \bbZ \comma b \in 2\bbZ) \comma \\
          \overline{ \TTheta_{34}(\tau)}^{4} 
       \cdot \chi^{E_8}_0(\tau)
       &
       (a \in 2 \bbZ \comma b \in 2\bbZ+1)  \comma \\
        \overline{ \TTheta_{23}(\tau)}^{4} 
       \cdot \chi^{E_8}_0(\tau)
       &
       (a \in 2 \bbZ+1 \comma b \in 2\bbZ) \comma \\
         -\overline{ \TTheta_{24}(\tau)}^{4} 
       \cdot \chi^{E_8}_0(\tau)      &
       (a \in 2 \bbZ+1 \comma b \in 2\bbZ+1) \comma
   \end{array}
   \right. 
   \label{E8cov}
\ee
satisfying the modular covariance (\ref{ModCov2}).
The  explicit form of the character
$
  \chi^{E_8}_0(\tau) 
$
is given  in (\ref{E8 ch}).
\subsection{Euclidean even self-dual lattices}
\label{subsec.ESDL}

Another class of the lattices for which the modular properties of the Euclidean lattice sum
in  (\ref{FiberTw}) are well controlled
is those associated with Euclidean even self-dual lattices.
Precisely, we start from a Lorentzian lattice $(p_L, p_R)$ which is 
specified by the basis $e^{*i}$ of a Euclidean even self-dual lattice.
The vielbein $e_i$ and the metric $G_{ij} $ are determined by (\ref{eeG}).
The matrix corresponding to the Cartan matrix 
is defined in this case by $C_{ij} = 2 G_{ij}$, which fixes the  background moduli  $E_{ij}$
by adopting the relations (\ref{EC}).
With this setting, after the T-duality twist we are left with the sum over the Euclidean
even self-dual lattice in  (\ref{FiberTw}).%
\footnote{
A typical example of Euclidean odd self-dual lattices is $\bbZ^n$, which is also unique for dimensions 
$n \leq 8$ (see e.g. \cite{King}). 
It is also realized as the $D_{n}$ lattice with the conjugacy classes 
$(0)$ and $(1)$. However, 
the partition function is not compatible with the modular covariance of the form  (\ref{ModCov}):
Starting with $ Z^{T^n}_{(0,1)} = \overline{\TTheta}_{34}^{n/2} \theta_3^n/\eta^n$,
and assuming the covariance  (\ref{ModCov}),
successive transformations $STSTS (\neq 1)$ would give 
$ Z^{T^n}_{(0,1)} = \overline{\TTheta}_{34}^{n/2} \theta_4^n/\eta^n$, in contradiction.
Choosing the basis so that 
$ e_i\cdot e_j = C_{ij}  =\delta_{ij}$, 
the integer-valuedness of $g_{SD}$
is not satisfied either if we adopt (\ref{EC}) and (\ref{gammaI}). 
} 

The Euclidean even self-dual lattices are allowed only for dimensions $d  \in 8 \bbZ$.
At $d=8$, the unique lattice is the $E_8$ lattice, which is already discussed in the previous
subsection. At $d=16$, there are two. One is the $E_8 \times E_8$ lattice and the other
is the Spin(32)/$\bbZ_2$ lattice.
At $d=24$, there are twenty four. These are called
Niemeier lattices. 

For an even-self dual lattice, $e^{*i}$ and $e_i$ span the same lattice since it is
self-dual. Moreover $e^{*i} \cdot e^{*j}, e_i\cdot e_j \in \bbZ$ ($i\neq j$) and 
they are even for $i=j$, since it is even.
Thus, $G^{ij}/2, E_{ij} \in \bbZ$ 
and the integer-valuedness of $g_{SD}$
for (\ref{gammaI}) is satisfied. 

The even self-duality also means that the modular property of the lattice sum
is trivial,
\be
  T^{ESD}_{\alpha\beta} = e^{-\frac{\pi d}{12}i} 
   \comma \qquad
      S^{ESD}_{\alpha\beta} =  1  
    \comma
\ee
as in the $E_8$ case.
Thus, denoting the corresponding character 
by $\chi^{ESD}$, 
the fiber partition function for a $d$-dimensional even self-dual lattice reads
\be
    Z^{T^d[ESD]}_{(a,b)}(\tau) = 
    \left\{
     \begin{array}{ll}
      \bigl{|}  \chi^{ESD}(\tau) \bigr{|}^2
       &
       (a \in 2 \bbZ \comma b \in 2\bbZ) \comma \\
          \overline{ \TTheta_{34}(\tau)}^{d/2} 
       \cdot \chi^{ESD}(\tau)
       &
       (a \in 2 \bbZ \comma b \in 2\bbZ+1) \comma  \\
        \overline{ \TTheta_{23}(\tau)}^{d/2} 
       \cdot \chi^{ESD}(\tau)
       &
       (a \in 2 \bbZ+1 \comma b \in 2\bbZ)  \comma \\
         (-1)^{d/8} \overline{ \TTheta_{24}(\tau)}^{d/2} 
       \cdot \chi^{ESD}(\tau)       &
       (a \in 2 \bbZ+1 \comma b \in 2\bbZ+1) \comma
   \end{array}
   \right.
\ee
satisfying the modular covariance (\ref{ModCov2}).
The explicit forms of $\chi^{ESD}$'s are found by using the relation of
these even-self dual lattices and Lie algebra lattices \cite{Lerche:1988np}.
For example
for $d=16$, the Spin(32)/$\bbZ_2$ lattice
is realized as the $D_{16}$ sublattices with
the vacuum and the spinor conjugacy class, and thus 
\be
\chi^{ESD}(\tau) =
   \chi^{{\rm Spin(32)}/\bbZ_2}(\tau)  
    = \chi_0^{D_{16}}(\tau) + \chi_s^{D_{16}} (\tau) \comma
\ee
in this case. Furthermore, by the identity of the Eisenstein series
$E_8(\tau)= E_4(\tau)^2$, one has $ \chi^{{\rm Spin(32)}/\bbZ_2} 
= \left[\chi^{E_8}_0(\tau)\right]^2 =  \chi^{E_8 \times E_8}$.

As in the case of the Englert-Neveu lattices, combining these with other parts,
we obtain the modular invariant
partition functions of the form (\ref{Ztot}) for the T-fold CFTs.
The action of $g_{SD}$ on $Z_{(a,b)}^{T^d[ESD]}$ in this case is $\bbZ_2$ both in the untwisted 
and twisted Hilbert spaces.

\subsection{Twists with phases}
\label{subsec.TwPh}

In acting with the T-duality transformation, the relative phase between the left and the right
mover are not  unique. Such a phase is strongly constrained 
when one requires that the full operator product expansion, not only the chiral one, 
of the vertex operators respects the invariance under the twist \cite{HW,Tan:2015nja}.
For the $A_1$ lattice, the phase in this case becomes $(-1)^{nw}$, with which 
the T-duality acts as an inner automorphism of $\mf{su}(2)_L \oplus \mf{su}(2)_R$
\cite{HW,KawaiS1}.
In section \ref{sec.hetero}, the possibility of including such phases is explicitly discussed,
when we apply our construction to the T-fold CFTs for the heterotic string.
In the partition function, the above  phase is implemented 
by the shift $\tau \to \tau+1/2$ since $p_L^2 - p_R^2 \in 2 \bbZ$.
It would  be an interesting problem if 
the phases in higher dimensional cases \cite{Tan:2015nja}
are also interpreted from the current algebra or the lattice point of view.

\section{Application to heterotic string theory}
\label{sec.hetero}

So far, we have discussed T-folds in bosonic string theory.
Our construction can be applied straightforwardly to the case of superstrings.
In particular, applying the results of the Englert-Neveu lattices for 
$D_2, D_4$ to type II superstrings
reduces to the analysis in \cite{Satoh:2015nlc}. Its generalization has also been
discussed \cite{Sugawara:2016lpa}. 
A notable point in these analyses is that, combined with further twists,
our T-fold CFTs simply realize the non-supersymmetric vacua with vanishing
cosmological constant at least at one loop.

In this section, we apply our construction to the heterotic string theory.
In our set up, the left-mover is the bosonic string with the $E_8\times E_8$-lattice, while the right-mover 
is the superstring including the fermionic one.
We focus on the supersymmetric models 
preserving 8 space-time supercharges.
Namely, we assume that the chiral reflection acts on the right-movers along a 
four dimensional fiber torus, 
which we choose to be $T^4[D_4]$ for simplicity. 
We 
briefly comment on the case $T^4[D_2 \times D_2] \equiv T^4[(A_1)^4]$ later on. 
The T-fold CFTs for the heterotic string have been discussed e.g. in 
\cite{Flournoy:2005xe,Bianchi:2012xz}.

However, since we are considering 
asymmetric orbifolds,
we still have a large variety of possibilities for the heterotic vacua:
the orbifold group  may act non-trivially
on {\em (i)} the left-mover of $T^4 {[} D_4 {]}$ and 
{\em (ii)} the 16-dim. internal torus with the $E_8\times E_8$ lattice,
while maintaining the modular invariance.
We demonstrate how we can systematically construct the modular invariants describing 
a large number of such heterotic string vacua,   
by utilizing the modular covariant blocks 
\eqn{A1cov}, \eqn{Drcov}, \eqn{E7cov} given in section 4. 
Above all, we 
uncover a fairly non-trivial phase factor that realizes the manifest modular covariance 
of the total building blocks.


\subsection{Orbifold action for heterotic T-folds}

Let us elaborate on a concrete construction of the models of heterotic T-folds. 
We start with the $E_8\times E_8$ heterotic string compactified on 
$T^4[D_4]$ ($X^{6, \ldots, 9}$-directions).
As in the previous section, we consider  
the orbifolding by $\sigma \equiv  \cT_{2\pi R} \otimes g$.
Here, $g$ acts along the $T^4$-direction as 
the chiral reflection,
\begin{equation}
g~ :~ 
 X^i_R ~ \longmapsto ~  -X^i_R,
~~~ \psi^i_R ~ \longmapsto - \psi^i_R,
~~~ (i=6,7,8,9),
\label{reflection}
\end{equation}
which preserves 1/2-SUSY,
while $\cT_{2\pi R}$ denotes the shift operator acting on the `base' $X^5$-direction,
\begin{equation}
\cT_{2\pi R} ~ :~ X^5\equiv X^5_L+ X^5_R ~ \longmapsto ~ X^5 + 2\pi R.
\label{shift}
\end{equation}
We use the notation $Z_{R,(w,m)}(\tau)$ defined in \eqn{S1Psum}
to write down the partition function for the $X^5$-direction.

We further allow $g$ to act non-trivially on the left-mover as the `chiral shifts' 
along various 
compact directions characterized by three integers $(r_1,r_2, r_3)$, 
where $r_1(\leq 4)$, $r_2,r_3 (\leq 8)$ are associated with $T^4[D_4]$ and the two $E_8$-directions.
Requiring the modular covariance, it 
turns out that this orbifold action 
provides  extra phases mentioned in section \ref{subsec.TwPh}.
We now separately specify the orbifold action $g$ on these sectors.

\paragraph{Action on  left-mover of  $T^4 [D_4 ] $-direction :}

~

As mentioned above, $g$ acts as the chiral reflection $(\-_R)^{\otimes 4}$  for the right-mover.

To specify the left-moving action, 
we consider the decomposition of the conjugacy
classes of $D_4$ for a fixed integer $r_1$ ($0 \leq r_1 \leq 4$),
\begin{align}
\La_{(\al)}^{D_4} = \sum_{\{\al_i\}, \beta } \, \left[ \La_{(\al_1)}^{A_1} 
\oplus \cdots \oplus \La_{(\al_{r_1})}^{A_1} \right]  \oplus \La_{(\beta)}^{X_{4-r_1}},
\label{decomp lattice D4}
\end{align} 
where $\al =0, v, s, c$ for $D_{4}$, and $\al = 0, 1$ for $A_1$
as in section \ref{subsec.LAL}. 
We also denote by $\al =0$ the conjugacy class for 
the basic representation (the root lattice itself)  
for any algebra $X$, i.e. $\Lambda_{(0)}^X \equiv \Lambda_X$.
We can uniquely determine the (semi-simple) Lie algebra
$X_{4-r_1}$ of rank $4-r_1$
on the R.H.S.
by imposing the following conditions;
\begin{description}
\item[(i)] $X_{4-r_1}$ is composed only of the irreducible components given in 
\eqn{SDG}, that is, $A_1$, $D_r$ ($r$: even), $E_7$, $E_8$.

\item[(ii)] $X_{4-r_1}$ is `maximal' in the following sense;
when taking  $\La^{D_4}_{(0)} \equiv \La_{D_4}$ 
on the L.H.S of \eqn{decomp lattice D4},
there is only one component with 
$\al_1=\cdots = \al_{r_1} =0$ on the R.H.S,
which should inevitably couple with $\La_{(0)}^{X_{4-r_1}} $.
(In other words, the conjugacy class $\La^{X_{4-r_1}}_{(\beta)}$ with $\beta \neq 0$
always couples with at least one spin 1/2-representation in the $r_1$ factors of $A_1$.)

\end{description}
We explicitly exhibit the solutions of $X_r$ in table \ref{table X}
for $ r \leq 4$. The entries for $r \geq 5$ in this table are used
shortly in the discussion on the $E_8$-part.

\begin{table}[t]
\begin{center}
\begin{tabular}{|c|c|c|c|c|c|c|c|c|c|}
\hline  
 $r$  & 0   & 1     & 2      & 3         & 4     & 5         & 6     & 7     &   8   \\ \hline
$X_r$ &  trivial   & $A_1$ & $D_2$  & $A_1 D_2$ & $D_4$ & $A_1 D_4$ & $D_6$ & $E_7$ & $E_8$ \\ \hline
\end{tabular}
\caption{list of $X_r$}
\label{table X}
\end{center}
\end{table} 


Then, we 
define the left-moving action of $g$ as an involution 
$g|_{\msc{left-mover}} = \left[ \rho_{A_1}\right]^{\otimes r_1}$ associated with the
lattice component  $[\La^{A_1}_{(*)}]^{r_1}$, where $\rho_{A_1}$ is an involutive (outer-)automorphism 
acting on the $\widehat{A}_1$-currents $\{J^a\}$ ($a=1,2,3$) as%
\footnote{
$\rho_{A_1}$ is explicitly written as 
$$
\rho_{A_1} = e^{-i\pi \frac{\ell}{2}} e^{i\pi J^3_0},
$$
on the integrable representation of spin $\ell/2$ ($\ell=0,1$). 
The phase factor $e^{-i\pi \frac{\ell}{2}}$ is necessary to make $\rho_{A_1}$ involutive. 
Note that the simpler inner-automorphism $\tilde{\rho}_{A_1} \equiv e^{i\pi J^3_0}$ is {\em not\/} involutive;
$$
\left[ \tilde{\rho}_{A_1} \right]^2 = (-1)^{\ell},
$$
which would play the role of the `$\bz_4$-chiral reflection' appearing in \cite{Satoh:2015nlc,Sugawara:2016lpa}.
It is presumably an interesting possibility to extend the heterotic vacua given in this section so as to include 
the $\bz_4$-action $\tilde{\rho}_{A_1}$, and we would like to discuss it elsewhere.
}
\begin{equation}
\rho_{A_1} J^a(z)   \rho_{A_1}^{-1} = -J^a(z) , ~~ (a=1,2), \hspace{1cm} 
\rho_{A_1} J^3(z)   \rho_{A_1}^{-1} = J^3(z). 
\label{rho A1}
\end{equation}
This operator is actually interpreted as the chiral half-shift along the direction of 
lattice $\La^{A_1}_{(*)}$ 
(up to some phase factor),
when  the $\widehat{A}_1$-currents $J^a$ are bosonized in the standard fashion.


We next consider the relevant partition sum with the orbifold twist $g$ inserted.
To this end, we recall that
$g$ acts on the right-mover as the chiral reflection $(\-_R)^{\otimes 4}$, which 
leaves the  sum  over the root lattice in the left-mover as in (\ref{FiberTw}).
Together with the condition {\bf (ii)} given above 
as well as the definition of $\rho_{A_1}$,
it is then obvious that only the basic representation of
 $\left[\widehat{A}_1\right]^{r_1} \oplus \widehat{X}_{4-r_1} $ can yield  non-vanishing 
contributions. 
The right-mover just  gives  
$\overline{ \left[ \TTheta_{34}(\tau)^{1/2} \right]^4} $, 
as already described in section 4.
On the other hand,
the $\left[ \rho_{A_1}\right]^{\otimes r_1}$-twist in the left-mover acts as sign factors 
on the relevant charge lattice, while leaving the oscillator parts unchanged, 
which again provides $\left[ \TTheta_{34}(\tau)^{1/2} \right]^{r_1} $
eventually. (See e.g. \cite{KawaiS1} for detail.)
In this way, we obtain
\begin{align}
\Tr_{T^4[D_4]}
\left[g q^{L_0-\frac{4}{24}} \overline{ q^{\tL_0-\frac{4}{24}} } \right] 
& 
= \overline{ \TTheta_{34}(\tau)^{2}} \cdot \TTheta_{34}(\tau)^{\frac{r_1}{2}}
\chi^{X_{4-r_1}}_{0}(\tau) 
\nn
&\equiv 
\left[ \overline{\tchi_{(0,1)}^{A_1}(\tau)} \right]^4 \cdot \left[\tchi_{(0,1)}^{A_1}(\tau)\right]^{r_1}
\chi^{X_{4-r_1}}_{(0,1)}(\tau).
\label{trace T4 g}
\end{align}
Here, the building blocks $\chi^{X_r}_{(0,1)}(\tau) \equiv \chi^{X_r}_{0}(\tau)$ 
from the lattice
and $\tchi^{A_1}_{(0,1)}(\tau) \equiv \TTheta_{34}(\tau)^{1/2}$ for the $\rho_{A_1}$-twist 
already appeared in \eqn{A1cov}, \eqn{Drcov}, \eqn{E7cov}, 
and are summarized in appendix \ref{block lattice} explicitly.  
For later convenience, we have also rewritten 
$\overline{\vartheta_{34}}^{1/2}$ in the right-mover
as $\overline{\tchi^{A_1}_{(0,1)}}$,
although it does not necessarily originate from the $\hat{A}_1$-symmetry.

For example, in the case of $r_1=1$, the relevant decomposition is 
\be
\La_{D_4} = \La_{(0)}^{A_1} \oplus \La_{(0)}^{X_{3}} + \cdots 
\equiv \La_{A_1} \oplus \La_{A_1} \oplus \La_{D_2} + \cdots .
\ee
and the trace \eqn{trace T4 g} becomes  
\be
\left[ \overline{\tchi_{(0,1)}^{A_1}(\tau)} \right]^4 \cdot \tchi_{(0,1)}^{A_1}(\tau)
\chi^{X_{3}}_{(0,1)}(\tau)\equiv 
\overline{\frac{\th_3^2 \th_4^2}{\eta^4}} \cdot \left[\frac{\th_3 \th_4}{\eta^2} \right]^{\frac{1}{2}} \chi^{A_1}_{0}(\tau) \chi^{D_2}_{0}(\tau),
\ee
where 
$ \chi^{A_1}_{0}(\tau)$, $\chi^{D_2}_{0}(\tau)$ are the characters of basic representations of $(\widehat{A}_1)_1$, $(\widehat{D}_2)_1$ respectively.


\paragraph{Action on two $E_8$-directions :}

~

Let us first focus on one of the $E_8$-factors. There, we have
a unique conjugacy class, that is, the root lattice itself.
We fix an integer $r_2$ ($0 \leq r_2 \leq 8$), 
and consider the decomposition of the 
root lattice $\La_{E_8}$ as 
\begin{align}
\La_{E_8} = \sum_{\{\al_i\}, \beta } \, \left[ \La_{(\al_1)}^{A_1} 
\oplus \cdots \oplus \La_{(\al_{r_2})}^{A_1} \right]  \oplus \La_{(\beta)}^{X_{8-r_2}}.
\label{decomp lattice E8}
\end{align}
The decomposition \eqn{decomp lattice E8} is again
uniquely determined by essentially the same conditions as for $T^4[D_4]$,
i.e. {\bf (i)} and {\bf (ii)} with $D_4, X_{4-r_1}$ replaced by $E_8, X_{8-r_1}$, respectively.
The result  of $X_{8-r_2}$ is listed in table \ref{table X}.%
\footnote
   {The uniqueness of the decomposition \eqn{decomp lattice E8} would be slightly non-trivial, 
   even though it is almost trivial for the $D_4$-case \eqn{decomp lattice D4}.
For instance, in the case of $r_2=1$, 
one might think that  
another decomposition 
$$
\La_{E_8} = \La_{(0)}^{A_1} \oplus \left( \La_{(0)}^{A_1} \oplus \La_{(0)}^{D_6} \right) + \La_{\msc{rem}},
$$ 
would be allowed.
However, 
$\La_{\msc{rem}}$ here includes the conjugacy class such as 
$
\La_{(0)}^{A_1} \oplus \left( \La_{(1)}^{A_1} \oplus \La_{(s)}^{D_6} \right).
$
Thus, this possibility is excluded by the condition {\bf (ii)}, 
and we obtain the unique decomposition with 
$X_7=E_7$.
}

We then define the $g$-action in this sector by
$\left[\rho_{A_1}\right]^{\otimes r_2}$ associated with the lattice component 
$[\La^{A_1}_{(*)}]^{r_2}$.
Since the relevant trace has contribution 
 only  from the basic representation of $X_{8-r_2}$, we have 
\begin{align}
\Tr_{E_8}
[g q^{L_0-\frac{8}{24}}  ] 
&=  
 \left[\tchi_{(0,1)}^{A_1}(\tau)\right]^{r_2}
\chi^{X_{8-r_2}}_{(0,1)}(\tau).
\label{trace E8 g}
\end{align}

The $g$-action for another $E_8$-factor is defined in the same way with an integer $r_3$ ($0\leq r_3 \leq 8$).


\subsection{Construction of heterotic T-folds}

Now, let us discuss how to construct the full building blocks characterized 
by the three integers $(r_1,r_2,r_3)$,
which are modular covariant.
In other words, we would like to construct the modular completions of \eqn{trace T4 g} and \eqn{trace E8 g}.
For this purpose we recall the modular covariant blocks 
\eqn{A1cov}, \eqn{Drcov}, \eqn{E7cov}, 
and consider their extensions to the $r$-dim. torus $T^r[X_r]$ composed of their products.
For the `odd sector' with $a \in 2\bz+1$ or $b \in 2\bz+1$, 
they are organized into the form,
\begin{equation}
Z^{T^r[X_r]}_{(a,b)}(\tau) := \ep^{[r]}_{(a,b)} \, \overline{\left[\tchi^{A_1}_{(a,b)}(\tau)\right]^r} \chi^{X_r}_{(a,b)}(\tau).
\label{Z Xr}
\end{equation}
Here $\ep^{[r]}_{(a,b)}$ denotes the phase factor assuring the modular covariance, which can be directly read off from 
\eqn{A1cov}, \eqn{Drcov}, \eqn{E7cov}, and 
generally expressed as 
\begin{equation}
\ep^{[r]}_{(a,b)} :=  e^{\frac{i\pi}{8} r (-1)^a ab} \left(\kappa_{(a,b)}\right)^r, \hspace{1cm} (a \in 2\bz+1 ~ \mbox{or} ~ b \in 2\bz +1),
\label{ep r}
\end{equation}
with 
\begin{equation}
\kappa_{(a,b)} := 
\left\{
\begin{array}{ll}
-1 & ~~~ a \equiv 3,5 ~ (\mod \, 8), ~ b \in 2\bz+1,
\\
1 & ~~~ \mbox{otherwise}.
\end{array}
\right.
\end{equation}
Note that the peculiar factor $\kappa_{(a,b)}$ 
affects only for odd $r$.

Furthermore, it is useful to note the following observations:
\begin{itemize}
\item The function $\left|\tchi^{A_1}_{(a,b)}(\tau)\right|^2$ 
satisfies the modular covariance of the form (\ref{ModCov}).

\item The function $(-1)^{ab} \, \left[ \tchi^{A_1}_{(a,b)}(\tau)\right]^8$ 
is similarly modular covariant up to the phases arising from the $T$-transformation acting on the factor $\eta(\tau)^{-8}$.

\end{itemize}
Based on these facts, one can 
construct the building blocks with the expected modular properties
as follows:
\begin{description}
\item [(1) $T^4{[}D_4{]}$-sector : ] 
Fix an integer $r$ ($0\leq r \leq 4 $), and set
\begin{align}
F^{[r]}_{(a,b)}(\tau) & := Z^{T^r[X_r]}_{(a,b)}(\tau) \cdot \left|\tchi_{(a,b)}^{A_1}(\tau)\right|^{2(4-r)}
\nn
& \equiv \ep^{[r]}_{(a,b)} \, \overline{\left[\tchi^{A_1}_{(a,b)}(\tau)\right]^4} \left[\tchi^{A_1}_{(a,b)}(\tau)\right]^{4-r} \chi^{X_r}_{(a,b)}(\tau).
\label{def Fab}
\end{align} 
By construction, $F^{[r]}_{(a,b)}(\tau)$ is obviously modular covariant as
\begin{equation}
\left. F^{[r]}_{(a,b)}(\tau) \right|_{T} =  F^{[r]}_{(a,a+b)}(\tau),
\hspace{1cm} 
\left. F^{[r]}_{(a,b)}(\tau) \right|_{S} =  F^{[r]}_{(b,-a)}(\tau).
\label{modular Fab}
\end{equation}

\item[(2) $E_8\times E_8$-sector : ]
For a single $E_8$-factor,
fix an integer $s$ ($0\leq s \leq 8 $), and define a chiral building block as 
\begin{align}
G^{[s]}_{(a,b)}(\tau) & :=  Z^{T^r[X_s]}_{(a,b)}(\tau) \cdot \left|\tchi_{(a,b)}^{A_1}(\tau)\right|^{-2 s} \cdot (-1)^{ab} \left[\tchi^{A_1}_{(a,b)}(\tau)\right]^8
\nn
& \equiv \ep^{[s]}_{(a,b)} (-1)^{ab} \, \left[\tchi^{A_1}_{(a,b)}(\tau)\right]^{8-s} \chi^{X_s}_{(a,b)}(\tau).
\label{def Gab}
\end{align}
Then, $G^{[s]}_{(a,b)}(\tau)$ is `almost' modular covariant, which precisely means the following modular properties,
\begin{equation}
\left. G^{[s]}_{(a,b)}(\tau) \right|_{T} =  e^{-2\pi i \frac{1}{3}}\, G^{[s]}_{(a,a+b)}(\tau),
\hspace{1cm} 
\left. G^{[s]}_{(a,b)}(\tau) \right|_{S} =  G^{[s]}_{(b,-a)}(\tau).
\label{modular Gab}
\end{equation}
Chiral blocks for another $E_8$-factor are identical.

\end{description}


To describe the total modular invariant, 
we still need to describe the free fermion chiral block in the right-mover, 
which is twisted by $(\-_{R})^{\otimes 4}$. 
This has been presented e.g. in \cite{Satoh:2015nlc,Sugawara:2016lpa}, and  
can be concisely expressed as 
\begin{align}
\overline{f_{(a,b)}(\tau)} =
& (-1)^{ab} \ep^{[4]}_{(a,b)} \, \overline{\left[\left(\tchi^{A_1}_{(a,b)}(\tau)\right)^4 -  \left(\tchi^{A_1}_{(a,b)}(\tau)\right)^4 \right]} ,
\quad 
 (a\in 2\bz+1 ~\mbox{or} ~ b \in 2\bz+1), \nn 
\label{def fab}
\end{align}
in terms of the notation adopted here. Here the trivial cancellation appearing in the bracket $[ \cdots ]$ just means the existence of supersymmetry.
A more explicit form of $f_{(a,b)}(\tau)$ is given in appendix \ref{blck fermion}.
The modularity of $\overline{f_{(a,b)}(\tau)}$ is expressed as 
\begin{equation}
\left. \overline{f_{(a,b)}(\tau)} \right|_{T} =  -e^{2\pi i \frac{1}{6}}\, \overline{f_{(a,a+b)}(\tau)},
\hspace{1cm} 
\left. \overline{f_{(a,b)}(\tau)} \right|_{S} =  \overline{f_{(b,-a)}(\tau)}.
\label{modular fab}
\end{equation}


Combining all the sectors, we can write down the total partition function characterized by three integers $(r_1,r_2,r_3)$ ($0\leq r_1 \leq 4$, $0\leq r_2 , r_3 \leq 8$) as follows:
\begin{align}
Z^{[r_1,r_2,r_3]}(\tau) 
 & = \frac{1}{2} Z^{5d}(\tau) \sum_{w,m\in \bz}\, Z_{R,(w,m)}(\tau) \, Z_{(w,m)}^{[r_1,r_2,r_3]}(\tau) ,
\label{Z total}
\end{align}
where 
\begin{align}
Z^{[r_1,r_2,r_3]}_{(w,m)}(\tau)
& : =  
\overline{\left[\left(\frac{\th_3}{\eta}\right)^4 - \left(\frac{\th_4}{\eta}\right)^4 - \left(\frac{\th_2}{\eta}\right)^4 \right]} \,
\sum_{j} 
\left| \chi^{D_4}_j(\tau) \right|^2 \left[\chi^{E_8}_0(\tau)\right]^2,
\label{SUSY model even}
\end{align}
for the `even sector' ($w,m\in 2\bz$), which is actually independent of $(r_1,r_2,r_3)$,  and 
\begin{align}
Z^{[r_1,r_2,r_3]}_{(w,m)}(\tau) 
 & : =  \overline{f_{(w,m)}(\tau)} F_{(w,m)}^{[4-r_1]}(\tau) 
 G_{(w,m)}^{[8-r_2]}(\tau) G_{(w,m)}^{[8-r_3]}(\tau)
\nn
& \equiv 
\ep^{[-\sum_i r_i]}_{(w,m)} \, 
\overline{\left[\left(\tchi^{A_1}_{(w,m)}(\tau)\right)^4 -  \left(\tchi^{A_1}_{(w,m)}(\tau)\right)^4 \right]} 
\overline{\left[ \tchi^{A_1}_{(w,m)}(\tau) \right]^{4}}
\nn
& \hspace{1cm}
\times
 \left[ \tchi^{A_1}_{(w,m)}(\tau) \right]^{\sum_i r_i}\, \chi_{(w,m)}^{X_{4-r_1}}(\tau)\,
\prod_{i=2,3} \chi_{(w,m)}^{X_{8- r_i}}(\tau) ,
\label{SUSY model odd}
\end{align}
for the `odd sector' ($w\in 2\bz+1$ or $m \in 2\bz+1$).
In \eqn{Z total}, we denote the (transverse part of) bosonic sector of the $X^{0, 1, \ldots, 4}$-directions as $Z_{\msc{5d}}(\tau)$, 
which is 
assumed to be modular invariant and not important here. 
The modular invariance of the total partition function  $Z^{[r_1,r_2,r_3]}(\tau) $ is now obvious due to the modular covariance of 
the building blocks $Z_{(w,m)}^{[r_1,r_2,r_3]}(\tau) $. Especially the covariance of the odd sector \eqn{SUSY model odd}
is readily confirmed by the relations \eqn{modular Fab}, \eqn{modular Gab} and \eqn{modular fab}.


We add a few comments:
\begin{itemize}
\item In the cases when all $r_i$ are even, only the $D_{2r}$-lattices
(or the $E_8$-lattice itself) 
come into the above construction. For these cases, our heterotic T-fold vacua can 
be reproduced by the free fermion construction. 
However, when  at least one of $r_i$ is odd,  our construction does not reduce to the free fermion construction.


\item It is straightforward to apply the above construction to the case of $T^4[D_2 \times D_2 ]$ 
$ \left( \equiv T^4[(A_1)^4]\right)$.
The lattice decomposition \eqn{decomp lattice D4} should be replaced with 
\begin{align}
\La_{(\al)}^{D_2 \times D_2} = \sum_{\{\al_i\}, \beta } \, \left[ \La_{(\al_1)}^{A_1} \oplus \cdots \oplus \La_{(\al_{r_1})}^{A_1} \right]  
\oplus \La_{(\beta)}^{Y_{4-r_1}},
\label{decomp lattice D2D2}
\end{align}
and the root system $Y_r$ $(0\leq r \leq 4)$ 
is uniquely determined as  for $T^4[D_4]$.
The result is explicitly listed in table \ref{table Y}.
Therefore, in order to construct the desired vacua, we only have  to replace $\chi_{(w,m)}^{X_{4-r_1}}(\tau)$ with $\chi_{(w,m)}^{Y_{4-r_1}}(\tau)$ 
in \eqn{SUSY model odd}, and also $\chi^{D_4}_j(\tau)$ with $\left[ \chi^{D_2}_j(\tau) \right]^2$ in \eqn{SUSY model even}. 

\end{itemize}

\begin{table}[t]
\begin{center}
\begin{tabular}{|c|c|c|c|c|c|}
\hline  
 $r$  & 0   & 1     & 2      & 3         & 4     \\ \hline
$Y_r$ &  trivial   & $A_1$ & $D_2$  & $A_1 D_2$ & $D_2 D_2$ \\ \hline
\end{tabular}
\caption{list of $Y_r$}
\label{table Y}
\end{center}
\end{table} 



\subsection{Unitarity in each winding sector}

In the heterotic string vacua we constructed above, the action of the orbifold twist 
$\sigma =\cT_{2\pi R}\otimes g $
is simple in the untwisted sector, namely, 
the unwound sector along $S^1_R$,
because $g$ is involutive on the untwisted Hilbert space, $g^2 = {\bf 1}$. 
However, the situation gets much more complicated in the twisted sectors, 
especially in the winding sectors with odd winding  
$w\in 2\bz+1$ due to the existence of the non-trivial phase factor 
$\ep^{[*]}_{(w,m)}$ given in \eqn{ep r}. 
It is thus not so obvious whether or not the string spectrum is unitary in each winding sector,
which is read off by the standard technique of the 
Poisson resummation with respect to the temporal winding $m$.
Namely, after summing over $m$ 
and rewriting the total partition function 
in the form, 
\begin{align}
& Z^{[r_1,r_2,r_3]}(\tau) = Z^{5d}(\tau)\,
\sum_{w\in \bz}\,\left[ Z^{(\sNS)}_w(\tau) + Z^{(\sR)}_w(\tau)
\right] \left(\equiv 0\right),
\label{totalZ}
\end{align}
each of $Z^{(\sNS)}_w(\tau)$ should be $q$- (and $\bar{q}$-) expanded
with coefficients belonging to $\bz_{\geq 0}$.

Although it would look more cumbersome because of the complexity 
of the phase factor $\ep^{[*]}_{(w,m)}$,
we can perform the Poisson resummation analysis in a manner following
\cite{SatohS,Satoh:2015nlc,Sugawara:2016lpa}.
After that, we can confirm that the above heterotic vacua are indeed unitary for an arbitrary choice of $(r_1,r_2,r_3)$.
We here briefly sketch how it works as follows:
\begin{itemize}
\item For the sectors with $w \in 2\bz$, it is easy to see the spectrum is unitary.
Indeed, 
since the fermion chiral block $\overline{f_{(a,b)}(\tau)}$ given in \eqn{def fab} (or \eqn{fab} for 
a more explicit form)
with $a \in 2\bz$, $b\in 2\bz+1$ 
vanishes because of the cancellation {\em only\/}  within the NS-sector,
we find
\be
Z_w^{(\sNS)}(\tau) = \left. Z^{(\sNS)}_{w}(\tau)\right|_{\msc{even} \, m} , 
~~~ \left. Z^{(\sNS)}_{w}(\tau)\right|_{\msc{odd} \, m} \equiv 0, \hspace{0.8cm} (\any w \in 2\bz),
\ee
where $Z^{(\sNS)}_{w}(\tau)|_{\msc{even} \, m} $
($Z^{(\sNS)}_{w}(\tau)|_{\msc{odd} \, m} $) denotes
the summation over the even (odd) temporal winding $m$.
The remaining $\left. Z^{(\sNS)}_{w}(\tau)\right|_{\msc{even} \, m} $ is then 
$q$-expanded in the desired form thanks to  the absence of 
the phase factor $\ep^{[*]}_{(w,m)}$ in \eqn{SUSY model even}. 

\item For the sectors with $w\in 2\bz+1$, 
we have 
\begin{align}
 Z^{(\sNS)}_{w}(\tau) 
& = \left. Z^{(\sNS)}_{w}(\tau)\right|_{\msc{even} \, m} 
+ \left. Z^{(\sNS)}_{w}(\tau)\right|_{\msc{odd} \, m} 
\nn
& = \left. Z^{(\sNS)}_{w}(\tau)\right|_{\msc{even} \, m} 
+ \left. Z^{(\sNS)}_{w}(\tau+1)\right|_{\msc{even} \, m} .
\label{eval Z odd w}
\end{align}
The equality in the second line follows just because of the covariance of total building blocks
under the modular $T$-transformation. 

\item Nextly, we evaluate  
$Z^{(\sNS)}_{w}(\tau)|_{\msc{even} \, m} $, 
$(w \in 2\bz+1)$
by using the Poisson resummation. 
The relevant computation is now straightforward, since the phase factor
 $\ep^{[*]}_{(w,m)}$ 
is relatively simple,
$
\ep^{[- \bar{r}]}_{(w,2m')} = e^{i\pi \frac{\bar{r}}{4}  w m'},
$
where we set $\bar{r} \equiv \sum_i r_i$.
Other types of phase factors may come from  $\chi^{X_r}_{(w,2m')}(\tau)$
as in \eqn{Dr ab}, \eqn{A1 ab} and \eqn{E7 ab}.
In any case, however, the relevant phase factors
always have the form such as $e^{2 \pi i \al m'}$
with some rational number $\al$. 
This yields the shift of the KK momentum,
$
\frac{1}{2R}n \, \rightarrow \frac{1}{2R}(n+\al),
$
and no extra phases are left. 
We thus obtain the $q$-expansion with positive coefficients belonging to $\frac{1}{2}\bz.%
$\footnote
  {The potential factor $1/2$ in the coefficients of $q$-expansion comes from the fact that the Poisson resummation is now made over $m \in 2\bz$ rather than $m\in \bz$.  }

\item Finally, we pick up the remaining sector,
$
Z^{(\sNS)}_{w}(\tau)|_{\msc{odd} \, m} \equiv \left. \left[ Z^{(\sNS)}_{w}(\tau)|_{\msc{even} \, m} \right] \right|_T.
$
As pointed out in \cite{SatohS}, this is Poisson resummed  into almost the same form as $Z^{(\sNS)}_{w}(\tau)|_{\msc{even} \, m} $, 
but with an extra minus sign in each term with the level mismatch $h-\tilde{h} \in \frac{1}{2} + \bz$.
In the end, we conclude that the total partition sum for the odd winding sector 
\eqn{eval Z odd w} is indeed  $q$-expanded 
only with the coefficients belonging to $\bz_{\geq 0}$.

\end{itemize}

\section{Conclusions}
We demonstrated that one can systematically construct the modular invariant 
partition functions for the T-fold CFTs by using the Lie algebra lattices.
We first discussed the case of bosonic strings. 
By the condition that the background moduli  is at a fixed point for 
a simple T-duality transformation realized as a chiral reflection, 
the possible Lie algebras for the Englert-Neveu lattices are restricted to the four cases listed in (\ref{SDG})
among simple simply-laced ones.
Based on  the fact that 
the characters of the level-one affine Lie algebras form a finite dimensional
representation of the modular group, the partition functions for the 
fiber torus part are found to satisfy the modular covariance  of the form (\ref{ModCov2}).
The results are listed in (\ref{A1cov}), (\ref{Drcov}), (\ref{E7cov}) and (\ref{E8cov}).
Together with the base part, 
summing up these gives the desired modular invariants for T-folds.
Similar constructions are possible also by using the Euclidean even self-dual lattices.

We then applied the above construction to the T-folds in the $E_8 \times E_8$ heterotic string theory.
As an example, we took a fiber torus representing the $D_4$ Englert-Neveu lattice.
Incorporating the non-trivial twists/phases in the left-moving sector, 
we obtained 
a class of modular invariant partition functions of the T-fold CFTs 
which are labeled by three integers.
In the twisted sectors, the partition functions in the left-mover 
are given by the building blocks obtained in the bosonic-string case, which 
are composed of the characters of the affine Lie algebras at level one.
After the Poisson resummation, one can also check the unitarity of the spectrum.
The case of the  $D_2 \times D_2$ torus was briefly discussed.

Our construction in the bosonic-string case formally resembles 
the truncation of the bosonic-string spectrum to the heterotic-string spectrum,
which is used to study the T-duality of the latter \cite{Giveon:1994fu,Giveon:1988tt}.
Indeed, one can start with a $(d+d')$-dimensional torus whose background moduli 
takes the same form as in the truncation,
\begin{equation}
E_{IJ} = 
 \begin{pmatrix}
   E_{ij}  & 0 \\
   A_{\mu j} & E_{\mu\nu} + \frac{1}{4} A_{\mu i} A_{\nu j} G^{ij} 
\end{pmatrix}
\hspace{1cm} ( i,j= 1, \ldots , d, ~~ \mu, \nu = 1, \ldots , d'),
\end{equation}
and proceed as in section \ref{sec.PfnTf} and \ref{sec.TLat}.
An interesting possibility in this case is that the additional moduli $A_{\mu k}$
may be incorporated in the T-fold CFTs. For this to be the case, 
one needs to check the fixed-point condition of the T-duality and 
 also to confirm that the twisted partition functions with non-trivial
$A_{\mu i}$ indeed satisfy the modular covariance of the form (\ref{ModCov2}).
We leave these as future problems.

It is worthwhile to remark that the heterotic T-folds we constructed 
include novel cases  
which contain rather non-trivial phase factors and are {\em not\/} reduced 
to the free fermion construction. 
It would thus be interesting to apply our construction to 
building the `realistic'  heterotic vacua of asymmetric orbifolds,
since recent attempts so far
are mainly based on the free fermion construction e.g. as in \cite{Bianchi:2012xz}
for the SUSY vacua and in 
\cite{Blaszczyk:2014qoa,Angelantonj:2014dia,Faraggi:2014eoa,Abel:2015oxa,Kounnas:2015yrc} 
for the SUSY-breaking ones.
Especially, it is indeed possible to extend the present construction to 
a variety of the non-SUSY heterotic T-folds by following \cite{Satoh:2015nlc,Sugawara:2016lpa}.
It would also be interesting to figure out the moduli space of such a  class of vacua. 
We would like to return to these issues 
in a future work.



~


\begin{center}
{\large\bf Acknowledgments}
\end{center}
We would like to thank C. Ahn and Z. Bajnok for useful comments.
This work is supported in part by JSPS Grant-in-Aid for Scientific Research
24540248 and Japan-Hungary Research Cooperative Program 
from Japan Society for the Promotion of Science (JSPS).
\par\bigskip




\appendix
\renewcommand{\theequation}{\Alph{section}.\arabic{equation}}

\vspace{5ex}


\section{Summary of building blocks}

In this appendix we summarize the definitions of the
building blocks that are repeatedly used in the main text.


\subsection
{Building blocks associated with Lie algebra lattices}
\label{block lattice}

As preparation, we summarize our conventions of theta functions and
the character formulas of affine Lie algebras relevant to our analysis.
\begin{description}

\item[Theta functions :  ] 

~
Our convention of theta functions are 
 \begin{align}
 & \dsp \th_1(\tau,z): =i\sum_{n=-\infty}^{\infty}(-1)^n q^{(n-1/2)^2/2} y^{n-1/2}
  \equiv  2 \sin(\pi z)q^{1/8}\prod_{m=1}^{\infty}
    (1-q^m)(1-yq^m)(1-y^{-1}q^m), \nn [-10pt]
   & \\[-5pt]
 & \dsp \th_2(\tau,z): =\sum_{n=-\infty}^{\infty} q^{(n-1/2)^2/2} y^{n-1/2}
  \equiv 2 \cos(\pi z)q^{1/8}\prod_{m=1}^{\infty}
    (1-q^m)(1+yq^m)(1+y^{-1}q^m), \\
 & \dsp \th_3(\tau,z): =\sum_{n=-\infty}^{\infty} q^{n^2/2} y^{n}
  \equiv \prod_{m=1}^{\infty}
    (1-q^m)(1+yq^{m-1/2})(1+y^{-1}q^{m-1/2}),  
\\
 &  \dsp \th_4(\tau,z): =\sum_{n=-\infty}^{\infty}(-1)^n q^{n^2/2} y^{n}
  \equiv \prod_{m=1}^{\infty}
    (1-q^m)(1-yq^{m-1/2})(1-y^{-1}q^{m-1/2}) ,
\\
& \eta(\tau)  : = q^{1/24}\prod_{n=1}^{\infty}(1-q^n),
 \end{align}
 where $q := e^{2\pi i \tau}$, $y:= e^{2\pi i z}$.
 We  use abbreviations, $\th_i (\tau) \equiv \th_i(\tau, 0)$ with $\th_1(\tau)\equiv 0$.
 %


\item[\bf Characters of $(\widehat{A}_1)_1$ : ]

~

The character of affine $A_1$ of level one ($(\widehat{A}_1)_1$) is 
written as 
\begin{equation}
\chi^{A_1}_0(\tau) 
:= \frac{\th_3(2\tau)}{\eta(\tau)}
~~~ (\mbox{basic rep.}),
\hspace{1cm} 
\chi^{A_1}_1(\tau) 
:= \frac{\th_2(2\tau)}{\eta(\tau)}  
~~~ (\mbox{spin $1/2$ rep.}).
\label{A1 ch}
\end{equation}
We also define
\begin{equation}
\chi^{A_1}_{\pm}(\tau) := \chi^{A_1}_0(\tau) \pm  \chi^{A_1}_1(\tau).
\end{equation}


\item[\bf Characters of $(\widehat{D}_r)_1$ $\, (r \in 2\bz_{>0})$ : ]

~

The characters of $(\widehat{D}_r)_1$ are given by
\begin{align}
&    \chi^{D_r}_0(\tau) :=  \frac{1}{2\eta(\tau)^r} \bigl[ \theta_3(\tau)^r + \theta_4(\tau)^r  \bigr] 
\hspace{1cm} (\mbox{basic rep.}),
\nn
& \chi^{D_r}_v(\tau) :=  \frac{1}{2\eta(\tau)^r} \bigl[ \theta_3(\tau)^r - \theta_4(\tau)^r  \bigr] 
\hspace{1cm} (\mbox{vector rep.}),
\nn
&
\chi^{D_r}_s(\tau) \equiv \chi^{D_r}_c (\tau) := \frac{\theta_2(\tau)^r}{2\eta(\tau)^r} 
\hspace{1cm}
(\mbox{spinor and cospinor rep.}).
\label{Dr ch}
\end{align}


\item[\bf Characters $(\widehat{E}_7)_1$ : ]

~

The character of $(\widehat{E}_7)_1$ is given by 
\begin{align}
\chi^{E_7}_0(\tau) 
& :=
\frac{1}{2\eta(\tau)^7} \left[\th_3(2\tau) \left(\th_3(\tau)^6+\th_4(\tau)^6\right) 
+ \th_2 (2 \tau) \th_2(\tau)^6\right],
\label{E7 ch 0}
\end{align}
for the basic representation, and by
\begin{align}
\chi^{E_7}_1(\tau) 
& :=
\frac{1}{2\eta(\tau)^7} \left[\th_2(2\tau) \left(\th_3(\tau)^6-\th_4(\tau)^6\right) + \th_3 (2 \tau) \th_2(\tau)^6\right],
\label{E7 ch 1}
\end{align}
for the fundamental representation ${\bf 56}$ with conformal weight $h=\frac{3}{4}$.
We also define
\begin{equation}
\chi^{E_7}_{\pm}(\tau) := \chi^{E_7}_0(\tau) \pm  \chi^{E_7}_1(\tau).
\end{equation}


\item[\bf Character of $(\widehat{E}_8)_1$ : ]

~

The root lattice of $E_8$ is the simplest example of even self-dual lattices, and the 
corresponding chiral block for strings
is the character of the basic representation of $(\widehat{E}_8)_1$,
\begin{equation}
\chi^{E_8}_0(\tau) :=
\frac{1}{2\eta(\tau)^8} \left[\th_3(\tau)^8+ \th_4(\tau)^8+ \th_2(\tau)^8\right].
\label{E8 ch}
\end{equation}


\end{description}

~


Now, 
we  summarize the functions used in order to compose the modular covariant blocks 
in the main text, which are associated with 
the Lie algebra lattices for $A_1, D_{r} (r \in 2 \bbZ_{>0}), E_7, E_8$:
\begin{align}
 \chi^{D_r}_{(a,b)}(\tau) 
& :=
\left\{
\begin{array}{ll}
\frac{1}{2 \eta(\tau)^r}
\left\{\th_3(\tau)^r + e^{\frac{i\pi r}{4} a} \th_4(\tau)^r\right\},
& ~~ (a\in 2\bz, ~ b\in 2\bz+1), \\
\frac{1}{2 \eta(\tau)^r}
\left\{\th_3(\tau)^r + e^{\frac{i\pi r}{4} b} \th_2(\tau)^r\right\},
& ~~ (a\in 2\bz+1, ~ b\in 2\bz), \\
\frac{1}{2 \eta(\tau)^r}
\left\{\th_4(\tau)^r + e^{\frac{i\pi r}{4} (a+b-1)} \th_2(\tau)^r\right\},
& ~~ (a\in 2\bz+1, ~ b\in 2\bz+1), \\
\end{array}
\right.
\label{Dr ab}
\\
 \chi^{A_1}_{(a,b)}(\tau) 
&:= 
\left\{
\begin{array}{ll}
\frac{1}{2}
\left\{
\chi^{A_1}_+(\tau) + e^{\frac{i\pi}{2} a} \chi^{A_1}_-(\tau)
\right\},
& ~~ (a\in 2\bz, ~ b\in 2\bz+1), \\
\frac{1}{\sqrt{2}}
\left\{
\chi^{A_1}_0 (\tau) + e^{\frac{i\pi}{2} b} \chi^{A_1}_1(\tau)
\right\},
& ~~ (a\in 2\bz+1, ~ b\in 2\bz), \\
\frac{1}{\sqrt{2}}
\left\{
\chi^{A_1}_0 (\tau) + e^{\frac{i\pi}{2}(a+ b-1) } \chi^{A_1}_1(\tau)
\right\},
& ~~ (a\in 2\bz+1, ~ b\in 2\bz+1), \\
\end{array}
\right.
\label{A1 ab}
\\
 \chi^{E_7}_{(a,b)}(\tau) 
&:= 
\left\{
\begin{array}{ll}
\frac{1}{2}
\left\{
\chi^{E_7}_+(\tau) + e^{-\frac{i\pi}{2} a} \chi^{E_7}_-(\tau)
\right\}
& ~~ (a\in 2\bz, ~ b\in 2\bz+1), \\
\frac{1}{\sqrt{2}}
\left\{
\chi^{E_7}_0 (\tau) + e^{-\frac{i\pi}{2} b} \chi^{E_7}_1(\tau)
\right\}
& ~~ (a\in 2\bz+1, ~ b\in 2\bz), \\
\frac{1}{\sqrt{2}}
\left\{
\chi^{E_7}_0 (\tau) + e^{-\frac{i\pi}{2}(a+ b-1) } \chi^{E_7}_1(\tau)
\right\}
& ~~ (a\in 2\bz+1, ~ b\in 2\bz+1), \\
\end{array}
\right.
\label{E7 ab}
\\
\chi^{E_8}_{(a,b)}(\tau)
& :=   (-1)^{ab} \chi^{E_8}_0(\tau), ~~~ (a \in 2\bz+1 ~ \mbox{or} ~ b\in 2\bz+1).
\label{E8 ab}
\end{align}
In each case, the integer labels $a$, $b$ 
characterize the spatial and temporal boundary conditions.
We also denote
\begin{align}
 \tchi^{A_1}_{(a,b)}(\tau) 
&:=
\left\{
\begin{array}{ll}
\dsp \frac{1}{\eta(\tau)} 
\sum_{n\in \bz} \, (-1)^n q^{n^2}
\equiv \sqrt{\frac{\th_3(\tau)\th_4(\tau)}{\eta(\tau)^2}},
& ~~ (a\in 2\bz, ~ b\in 2\bz+1),\\
\dsp \frac{\sqrt{2}}{\eta(\tau)} 
\sum_{n\in \bz} \, q^{\left(n+\frac{1}{4}\right)^2}
\equiv \sqrt{\frac{\th_3(\tau)\th_2(\tau)}{\eta(\tau)^2}},
& ~~ (a\in 2\bz+1, ~ b\in 2\bz),\\
\dsp 
\frac{\sqrt{2}}{\eta(\tau)} \sum_{n\in \bz} \,  (-1)^n q^{\left(n+\frac{1}{4}\right)^2}
\equiv \sqrt{\frac{\th_4(\tau)\th_2(\tau)}{\eta(\tau)^2}},
& ~~ (a\in 2\bz+1, ~ b\in 2\bz+1), \\
\end{array}
\right.
\label{tA1 ab}
\end{align}
for the $(\widehat{A}_1)_1$-characters twisted by 
the involution $\rho_{A_1} \equiv e^{-\frac{ i \pi}{2} \ell} e^{i\pi J^3_0}$ in \eqn{rho A1}.
See e.g. appendix C of 
\cite{KawaiS1} for more detail. 
These coincide with $\TTheta_{pq}^{1/2}$ defined in (\ref{thetapq}).
It is easy to confirm the equality
$$
\chi^{D_2}_{(a,b)}(\tau) = \left[ \chi^{A_1}_{(a,b)}(\tau) \right]^2,
$$
which is consistent with the isomorphism of the root lattices
$\La_{D_2} \cong \La_{A_1} \oplus \La_{A_1}$.
We also use the notation such as 
$\chi_{(a,b)}^{X_3}(\tau) \equiv \chi_{(a,b)}^{A_1}(\tau) \chi_{(a,b)}^{D_2}(\tau)$ for $X_3 = A_1 D_2$ 
(see table \ref{table X}).


\subsection
{Free fermion chiral blocks}
\label{blck fermion}

We next describe the chiral blocks for the 8 world-sheet fermions $\psi_L^i$ twisted by 
\be
({\bf -1}_L)^{\otimes 4} ~ : ~ \psi^i_L ~ \longmapsto ~ -\psi^i_L, \hspace{1cm} (i=6, \ldots, 9).
\ee 
The relevant blocks are explicitly written as 
\begin{eqnarray}
 f_{(a,b)}(\tau)
&:=  & 2 q^{\frac{1}{4}a^2}e^{\frac{i\pi}{2}ab}
\,
\left(\frac{\th_1\left(\tau,\frac{a\tau+b}{2}\right)}{\eta(\tau)}\right)^2
\left(\frac{\th_1(\tau,0)}{\eta(\tau)}\right)^2
\nn
& \equiv &
(-1)^{ab} e^{\frac{i\pi}{2}ab}  \left[\left(\tchi^{A_1}_{(a,b)}(\tau)\right)^4 -  \left(\tchi^{A_1}_{(a,b)}(\tau)\right)^4 \right]
\nn
&\equiv &
\left\{
\begin{array}{ll}
 e^{\frac{i\pi}{2}ab}
\left\{
\left(\frac{\th_3}{\eta}\right)^2\left(\frac{\th_4}{\eta}\right)^2
- \left(\frac{\th_4}{\eta}\right)^2\left(\frac{\th_3}{\eta}\right)^2
+0 \right\}
 &  ~~ (a\in 2\bz,~ b\in 2\bz+1)\\
 e^{\frac{i\pi}{2}ab}
\left\{
\left(\frac{\th_3}{\eta}\right)^2\left(\frac{\th_2}{\eta}\right)^2 +0 
- \left(\frac{\th_2}{\eta}\right)^2\left(\frac{\th_3}{\eta}\right)^2
\right\}
 &  ~~ (a\in 2\bz+1,~ b\in 2\bz)\\
-  e^{\frac{i\pi}{2}ab}
\left\{ 0+
\left(\frac{\th_2}{\eta}\right)^2\left(\frac{\th_4}{\eta}\right)^2
- \left(\frac{\th_4}{\eta}\right)^2\left(\frac{\th_2}{\eta}\right)^2
\right\}
 &  ~~ (a\in 2\bz+1,~ b\in 2\bz+1).
\end{array}
\right.
\nn
&&
\label{fab}
\end{eqnarray}
In the last line, each term corresponds to the NS, $\tNS$, R sectors with  keeping this order.
These trivially vanish, as is consistent with the space-time SUSY. 
They satisfy the modular covariance of the form,
\begin{eqnarray}
 && f_{(a,b)}(\tau)|_S \equiv f_{(a,b)}\left(-\frac{1}{\tau}\right)
= f_{(b,-a)}(\tau), \nn
&& f_{(a,b)}(\tau)|_T \equiv f_{(a,b)}(\tau+1)
= - e^{-2\pi i \frac{1}{6}} f_{(a,a+b)}(\tau) .
\label{mc fab}
\end{eqnarray}
Therefore,
setting 
\be
\cJ(\tau) := \left(\frac{\th_3}{\eta}\right)^4
- \left(\frac{\th_4}{\eta}\right)^4 - \left(\frac{\th_2}{\eta}\right)^4, 
\ee
we find 
\begin{eqnarray}
 && \left. \left[ \overline{\cJ(\tau)}f_{(a,b)}(\tau)\right] \right|_S 
\equiv 
\overline{\cJ\left(-\frac{1}{\tau}\right)}
f_{(a,b)}\left(-\frac{1}{\tau}\right)
= \overline{\cJ(\tau)}f_{(b,-a)}(\tau), \nn
&& 
\left. \left[
\overline{\cJ(\tau)}
f_{(a,b)}(\tau)\right] \right|_T \equiv 
\overline{\cJ(\tau+1)}
f_{(a,b)}(\tau+1)
= \overline{\cJ(\tau)} f_{(a,a+b)}(\tau) .
\label{mc fab refined}
\end{eqnarray}
Since \eqn{fab}  vanishes, the relations \eqn{mc fab} may appear to be ambiguous. 
 See, however,  \cite{Satoh:2015nlc} for more precise arguments.


\newpage 
\setlength{\parskip}{0ex}

\end{document}